\documentclass[preprint,pre,amsmath]{revtex4-2}

\usepackage[pdftex]{graphicx} 
\usepackage{color}
\usepackage{bm}
\usepackage{epsfig}
\usepackage{latexsym}
\usepackage{amsmath}
\usepackage{nameref,hyperref}
\usepackage{cases,setspace,amsmath}
\usepackage{wasysym}

\makeatletter
\let\old@makecaption=\@makecaption
\usepackage{subcaption}
\let\@makecaption=\old@makecaption
\makeatother
\usepackage{blindtext}

\def\be{\begin{equation}}
\def\ee{\end{equation}}
\def\bea{\begin{eqnarray}}
\def\eea{\end{eqnarray}}
\def\bsn{\begin{subnumcases}}
\def\esn{\end{subnumcases}}

\begin{document}

\title[Finite-time quenches]{Kibble-Zurek scalings and coarsening laws in slowly quenched classical Ising chains}

\author{Lakshita Jindal}
                               \email{lakshita@jncasr.ac.in}
                                \author{Kavita Jain} 
                                \email{jain@jncasr.ac.in}
\affiliation{Theoretical Sciences Unit, \\
Jawaharlal Nehru Centre for Advanced Scientific Research, \\Bangalore 560064, India}

\date{\today}

\clearpage

\begin{abstract}
We consider a one-dimensional classical ferromagnetic Ising model when it is quenched from a low temperature to zero temperature in finite time using Glauber or Kawasaki dynamics. Most of the previous work on finite-time quenches assume  that the system is initially in equilibrium and focus on the excess defect density at the end of the quench which decays algebraically in quench time with Kibble-Zurek exponent. Here we are interested in understanding the conditions under which the Kibble-Zurek scalings do not hold and in elucidating the full dynamics of the defect density.  
 We find that depending on the initial conditions and quench time, the dynamics of the defect density can be characterized by coarsening and/or the standard finite-time quench dynamics involving adiabatic evolution and Kibble-Zurek dynamics; the time scales for crossover between these dynamical phases are determined by coarsening time and stationary state relaxation time. As a consequence, the defect density at the end of the quench is either a constant or decays following coarsening laws or Kibble-Zurek scaling. For the Glauber chain, we formulate a low temperature scaling theory and find exact expressions for the final defect density for various initial conditions. For the Kawasaki chain where the dynamic exponents for coarsening and stationary state dynamics are different, we verify the above findings numerically  and also examine the effect of unequal dynamic exponents.
\end{abstract}
 
 \maketitle
 
 \clearpage

\section{Introduction}
\label{intro}

The phenomenon of phase ordering in systems exhibiting thermal phase transition between a disordered and an ordered phase has received considerable attention in the past few decades \cite{bray2002theory}. Following a sudden quench from a disordered phase to a low temperature symmetry-broken phase, the system does not order instantaneously; instead, the domains of symmetry-broken phase grow locally until the system reaches equilibrium at large times \cite{bray2002theory}. 
Even if the system is quenched at a finite rate, it is not in equilibrium at the end of the quench and there are more defects compared to the equilibrium state at the quench temperature; one can then ask how the excess defect density decays with the quench time and what the dynamics are while the system is being slowly cooled.  

Some of these questions can be addressed in the framework of a theory first proposed by Kibble to describe the symmetry-breaking in the early universe \cite{kibble1976topology, kibble1980some} and later extended by Zurek to condensed matter systems \cite{zurek1985cosmological,zurek1996cosmological}.  For a classical or quantum system that shows a continuous phase transition between a disordered and an ordered phase, under a time-dependent change of a control parameter such as temperature, the Kibble-Zurek (KZ) theory predicts that there are more defects at the end of the slow quench than in equilibrium and the residual density of defects, in general,  decays algebraically in quench time with an exponent that depends on equilibrium critical exponents and quench protocol.  

These results can be understood by noting that if the system is initially deep in the high-symmetry phase,  there is a competition between two timescales, {\it viz.}, the equilibrium relaxation time  and the time remaining until the end of the quench. Away from the critical region where the correlation length is small, as the equilibrium relaxation time is much smaller than the time left until  the end of the quench, the system has sufficient time to relax in response to the changing temperature and reaches the equilibrium state at the instantaneous temperature ({\it adiabatic phase}). 
However, in the critical region, as the equilibrium correlation length and hence the relaxation time diverges, the system is unable to relax and falls out of equilibrium ({\it KZ phase}) so that there are more defects than had the system been in equilibrium. Assuming that the dynamics are `frozen' in the nonequilibrium phase, the KZ argument predicts how the excess defect density scales with the quench time \cite{kibble1980some,zurek1996cosmological}.

The slow annealing problem described above has been studied theoretically in various condensed matter systems including  classical \cite{reiss1980structures,schilling1988slow, cornell1991domain,brey1994dynamical,aliev2009generating, biroli2010kibble, krapivsky2010slow, liu2014dynamic, jain2016critical,ricateau2018critical, jeong2019growth, jeong2020nonequilibrium, Priyanka_2021, mayo2021distribution, kim2022nonequilibrium, godreche2022glauber} and quantum  \cite{polkovnikov2005universal, dziarmaga2010dynamics, dutta2015quantum, chandran2012kibble, chandran2013kibble, chesler2015defect} Ising models, and the KZ predictions have also been verified in  experiments \cite{del2014universality} on a wide variety of systems such as non-Newtonian fluids \cite{casado2006testing}, colloidal monolayers \cite{deutschlander2015kibble} and ultra-cold atomic gases \cite{beugnon2017exploring}. However, it has also been shown in theoretical studies that if the system is quenched deep in the ordered phase, the excess defect density follows coarsening (and not the KZ) scaling laws \cite{biroli2010kibble,jain2016critical}, and that the dynamics are not frozen in the KZ phase which affects the amplitude (but not the KZ exponent) of the defect density at the end of the quench \cite{Priyanka_2021}.

The KZ argument described above assumes that the system is initially in equilibrium state and far from the critical region. However, if the system starts in a nonequilibrium state or it is initially equilibrated to a temperature in the critical region and the quench time is not long enough for the system to reach the adiabatic phase, the KZ scalings may not hold. 
For the Ising chains studied here, we find that if the quench time is small compared to the equilibrium relaxation time, the system initially equilibrated to a temperature in the critical region can not relax to the perturbations arising due to changing temperature and the defect density at the end of the quench remains close to its initial value; on the other hand, if the system is initially not in equilibrium and quench times are small relative to the critical coarsening time scales, the defect density decreases following critical coarsening laws. For larger quench time (relative to the appropriate relaxation time), however, we find that the KZ scalings hold provided the system size is large enough. 
Interestingly, for the infinitely long Glauber chain, we show that the defect density at the end of quench can be captured exactly by a single expression [see, (\ref{e14}), (\ref{neqD2}), (\ref{neqD1}) for different initial conditions] for both small and large quench times. 

In the following sections, we study these scenarios in detail; besides, the results mentioned above for the defect density at the end of the quench, we also obtain analytical expressions for its temporal evolution in the Glauber Ising chain. We find that depending on the initial state and quench time, the dynamical evolution can be characterized by coarsening and/or the standard adiabatic evolution and KZ dynamics, and the crossover between these dynamical phases occurs on time scales that depend on the nonequilibrium and equilibrium relaxation times. Some of these results are also verified numerically for the Kawasaki chain for which, unlike the Glauber model, the dynamic exponent characterizing coarsening phenomenon is different from the stationary state dynamic exponent.


\section{Glauber Ising chain}


\subsection{Model}
\label{GMo}

We consider a one-dimensional ferromagnetic Ising model with nearest neighbor interactions defined by the Hamiltonian
 \begin{equation}
     H = -\sum_{i=1}^{L} \sigma_{i}\sigma_{i+1}
     \label{equn0}
    \end{equation}
 where the spin variable, $\sigma_i = \pm 1$ at site $i$ and $\sigma_{L+1}=\sigma_1$ as we assume periodic boundary condition for a 
 finite-sized system. In the equilibrium state, the correlation length $\xi_{eq}(T) \sim e^{\frac{2}{T}}$ diverges at the critical temperature $T=0$.

 To study the finite-time quench dynamics, in this section, we consider the Glauber dynamics in which the system evolves via single spin-flip and the total magnetization is not conserved \cite{glauber1963time}. The probability that the system is in a configuration $\{ \sigma_{1},\sigma_{2},\ldots,\sigma_{L} \}$ at time $t$ is described by 
 \begin{equation}
 \begin{split}
     \frac{d}{dt} p(\sigma_{1},\ldots,\sigma_{i},\ldots,\sigma_{L},t) = \sum_{i=1}^L \Big[w(-\sigma_{i} \to \sigma_i,t)\; p(\sigma_{1},\ldots,
     -\sigma_{i},\ldots,\sigma_{L},t)- \\ w(\sigma_{i} \to -\sigma_i,t)\;p(\sigma_{1},\ldots,\sigma_{i},\ldots,\sigma_{L},t) \Big]  
     \label{equ0}
 \end{split}
 \end{equation}
where $w(\sigma_{i} \to -\sigma_{i} ,t)$ is the transition rate at which the spin $i$ flips at time $t$ and is given by \cite{glauber1963time, Priyanka_2021}
 \begin{equation}
     w(\sigma_i \to -\sigma_{i},t) =1-\frac{\gamma(t)}{2}\sigma_i(\sigma_{i-1} + \sigma_{i+1})
     \label{transprob}
 \end{equation}
 with
 \be
 \gamma(t) = \textrm{tanh}\left( \frac{2}{T(t)} \right) \label{gamdef}
 \ee 

Previous work \cite{krapivsky2010slow, Priyanka_2021} have shown that if the system is cooled faster than a logarithmic decay in time (for example, if the temperature is decreased algebraically in time), the dynamics are essentially the same as that for infinitely rapid quench (up to logarithmic factors, see (23) and (24) of \cite{Priyanka_2021}).  We therefore consider the cooling protocol, $T(t) \sim -4\left[\ln \{(1-\gamma_0)(1-\frac{t}{\tau})^\alpha \} \right]^{-1}$, or 
\begin{equation}
    1- \gamma(t) = (1-\gamma_0)\Big(1-\frac{t}{\tau}\Big)^\alpha, \hspace{1cm} {\alpha>0} 
    \label{equn2}
\end{equation}
which states that the system is initially at a temperature $T_0$ where $\gamma_0 = \textrm{tanh}(2/T_0)$ and then cooled to zero temperature in a finite time $\tau$. Note that the above finite-time quench protocol reduces to instantaneous quench problem when the parameter $\alpha \to \infty$.

\subsection{Dynamics of spin-spin correlation function}

In the following, we study the equal time spin-spin correlation function 
\be
G_k(t)=\langle\sigma_i(t)\sigma_{i+k}(t)\rangle
\label{equaltime}
\ee
where the angular brackets denote the average with respect to the distribution $p(\{\sigma_i\},t)$. One can write down the time-evolution equation for $G_k(t)$ by multiplying both sides of (\ref{equ0}) with $\sigma_i\sigma_{i+k}$ and summing over all the possible configurations \cite{glauber1963time}. For time-dependent $\gamma$, we then obtain \cite{reiss1980structures,brey1994dynamical,krapivsky2010slow,Priyanka_2021}
 \begin{equation}
     \frac{d G_k}{dt} = -2G_k + \gamma(t) (G_{k-1} + G_{k+1}), \hspace{1cm}  k=1,...,L-1
     \label{equn1}
 \end{equation}
 with the boundary conditions, $G_0(t) = G_L(t) = 1$ and the initial condition, $G_k(0)$. The mean domain wall density is related to the correlation function as
 \be
 D(t)=\frac{1-G_1(t)}{2} \label{ddendef}
 \ee

We first briefly summarize the known results that are pertinent to the discussion in the following subsections. When an infinitely large system is instantaneously cooled (or heated) to a time-independent temperature $T$, the general solution for the two-point correlation function is given by (63) of \cite{glauber1963time}: 
\be
G_k(t) = G_{k,eq} + e^{-2t}\sum_{l=1}^\infty (G_l(0) - G_{l,eq}) [I_{k-l}(2\gamma t) - I_{k+l}(2\gamma t)] \label{eq3}
\ee
where $\gamma(T)$ is given by (\ref{gamdef}) for a constant temperature $T$ and $I_{\nu}(z)$ is the modified Bessel function of the first kind. At $t \to \infty$, the system reaches the equilibrium state at temperature $T$ where the correlation function is given by (54) and (56) of \cite{glauber1963time}: 
\begin{equation}
G_{k,eq}=\Bigg(\frac{1-\sqrt{1-\gamma^2}}{\gamma}\Bigg)^k
\label{equ9}
 \end{equation}
 from which we obtain the equilibrium defect density at a low temperature $T \ll 1$ to be 
\be
 D_{eq}(T) \approx \sqrt{\frac{1-\gamma}{2}} \label{equildef} ~,~ \gamma \to 1
 \ee
As shown in Appendix~\ref{cnstT}, if the system is instantaneously quenched from a high temperature to a low temperature $T$, the defect density at large times decays as \cite{glauber1963time} 
 \be
 D(t) \approx \frac{1}{2\sqrt{\pi t}} \left[1+2 (1-\gamma)t \right]~,~1 \ll t \ll (1-\gamma)^{-1}
 \label{defs}
 \ee
 Likewise, if the system is initially in the equilibrium state at zero temperature and is instantaneously heated to $0 < T \ll 1$, the defects in the system increase as $\sqrt{t}$ (see Appendix~\ref{cnstT}): 
 \be
 D(t) \approx 2(1-\gamma)  \sqrt{\frac{t}{\pi}}~,~ 1 \ll t \ll (1-\gamma)^{-1}
\label{defc}
 \ee

On the other hand, if an infinitely large system is cooled to zero temperature in a finite time $\tau$ using an arbitrary protocol $\gamma(t)$, an exact expression for $G_k(t)$ can be written as  \cite{brey1994dynamical,Priyanka_2021}  
 \bea
G_{k,eq}(t) &-& G_k(t)= \int_0^t dt' \;e^{-2(t-t')}\;\frac{d\gamma(t')}{dt'}\int_0^\pi \frac{dq}{\pi} \frac{\textrm{sin}(kq)\;\textrm{sin}(q)}{(1-\gamma(t')\;\textrm{cos}(q))^2}\;e^{2\;\textrm{cos}(q)\int_{t'}^t dy \; \gamma(y)} \nonumber \\
&+& \frac{2}{\pi}\int_0^{\pi} dq\;\sin({kq})\;e^{-2\int_0^t dy(1-\gamma(y)\cos({q}))} \sum_{m=1}^{\infty} \sin({mq})\;(G_{m,eq}(0)-G_m(0)) 
\label{Gksoln}
\eea
Using the above result, the dynamics of the spin-spin correlation function in the KZ phase have been studied in detail for the cooling protocol (\ref{equn2}) starting from an infinite temperature ($\gamma_0 = 0$), and it is shown that for large $\tau$, the mean defect density at the end of the quench is given exactly by \cite{Priyanka_2021}
\begin{equation}
D(\tau)= \frac{1}{2\sqrt{\pi}}\Big(\frac{2}{1+\alpha}\Big)^{\frac{1}{2(1+\alpha)}}\Gamma\Big(\frac{1+2\alpha}{2+2\alpha}\Big) \times \frac{1}{\tau^{\frac{\alpha}{2(1+\alpha)}}}
\label{eq1}
\end{equation}
which, for a given $\tau$, decreases monotonically with exponent $\alpha$. For $\alpha \to 0$, the system stays close to the initial high temperature until $\tau$ and therefore carries almost all the initial defects till the end of the quench, while  for $\alpha \to \infty$, the number of defects decrease via coarsening and (\ref{eq1}) coincides with (\ref{defs}) when $\gamma=1$.

The $\tau$-scaling in (\ref{eq1}) can be understood using the Kibble-Zurek argument \cite{kibble1980some,zurek1996cosmological}: below a time scale $\hat{t}$, the system can relax to the equilibrium state at the instantaneous temperature ({\it adiabatic phase}: $t \ll {\hat t}$) and therefore, the relevant time scale in this regime is the equilibrium relaxation time, $\xi^{z_{eq}}_{eq}(t) \sim e^{\frac{2 z_{eq}}{T(t)}} \sim (1-\gamma(t))^{-\frac{z_{eq}}{2}} \ll {\hat t}$. But above $\hat{t}$ where the system can not relax due to diverging correlation length ({\it KZ phase}: ${\hat t} \ll t < \tau$), the only time scale is the time remaining until the quench ends, {\it viz.}, $\tau-{\hat t}$. 
For the cooling protocol (\ref{equn2}), these two time scales are comparable when 
 \be
  \tau - \hat{t} \sim (1-\gamma_0)^{\frac{-z_{eq}}{2+\alpha z_{eq}}}\;\tau^{\frac{\alpha z_{eq}}{2 + \alpha z_{eq}}}
  \label{tscale}
 \ee
 Assuming that the dynamics during ${\hat t} < t < \tau$ can be neglected, the mean defect density $D(\tau) \sim D({\hat t}) \sim \xi^{-1}_{eq}({\hat t})$ and yields  
 \be
 D(\tau) \sim (1-\gamma_0)^{\frac{1}{2+\alpha z_{eq}}}\;\tau^{-\beta} \label{KZIsing}
 \ee
 where the KZ exponent 
 \be
 \beta=\frac{\alpha}{2 + \alpha z_{eq}} \label{betadefn}
 \ee
 which matches the $\tau$-dependence in (\ref{eq1}) on using that the stationary state dynamic exponent $z_{eq}=2$ \cite{glauber1963time,cordery1981physics} for the Glauber chain. 
 
The expression for the spin-spin correlation function given in (\ref{Gksoln}) is valid for infinitely large system and arbitrary initial temperature $T_0$, and has been analyzed for large $\tau$ and high $T_0$ \cite{Priyanka_2021}.
Here we are interested in the scenario when $T_0 \ll 1$. But as the double integrals appearing in  (\ref{Gksoln}) are quite involved, in Appendix~\ref{smalleqG}, we develop a scaling theory for low initial temperature or large initial correlation length. For infinitely long chain, we find that for $\tau \to \infty, \gamma_0 \to 1$ with finite $\tau (1-\gamma_0)$, the spin-spin correlation function is given by
\bea
G(y,x) &=& \frac{2}{\pi}\int_{0}^{\infty}dq\;\sin(qy)\;q\; e^{-q^2 x}\int_{0}^{x}dw\;e^{q^2 w -\frac{2\lambda_0}{\alpha+1}\big((1-w)^{\alpha+1}-(1-x)^{\alpha+1}\big)} \nonumber \\ 
&+& \sqrt{\frac{2}{\pi}}\int_{0}^{\infty}dq\;\sin(qy)\;\tilde{G}(q,0)\;e^{-q^2 x +\frac{2\lambda_0}{\alpha+1}\big((1-x)^{\alpha+1}-1\big)}
\label{Gyx}
\eea
where
\be
y=\frac{k}{\sqrt{\tau}}, x=\frac{t}{\tau}, \lambda_0=\tau (1-\gamma_0)
\label{scaledefn}
\ee
and $\tilde{G}(q,0)$ is the sine transform of the initial condition $G(y,0)$. Furthermore, the effect of finite system size is discussed using a scaling argument. We also compare our analytical results with the numerical  solution of the exact equation (\ref{equn1}).

\subsubsection{Equilibrium initial condition}
\label{EIC}

\begin{figure}[t]
       \begin{subfigure}{0.49\textwidth}
         \centering
         \includegraphics[width=1.0\textwidth]{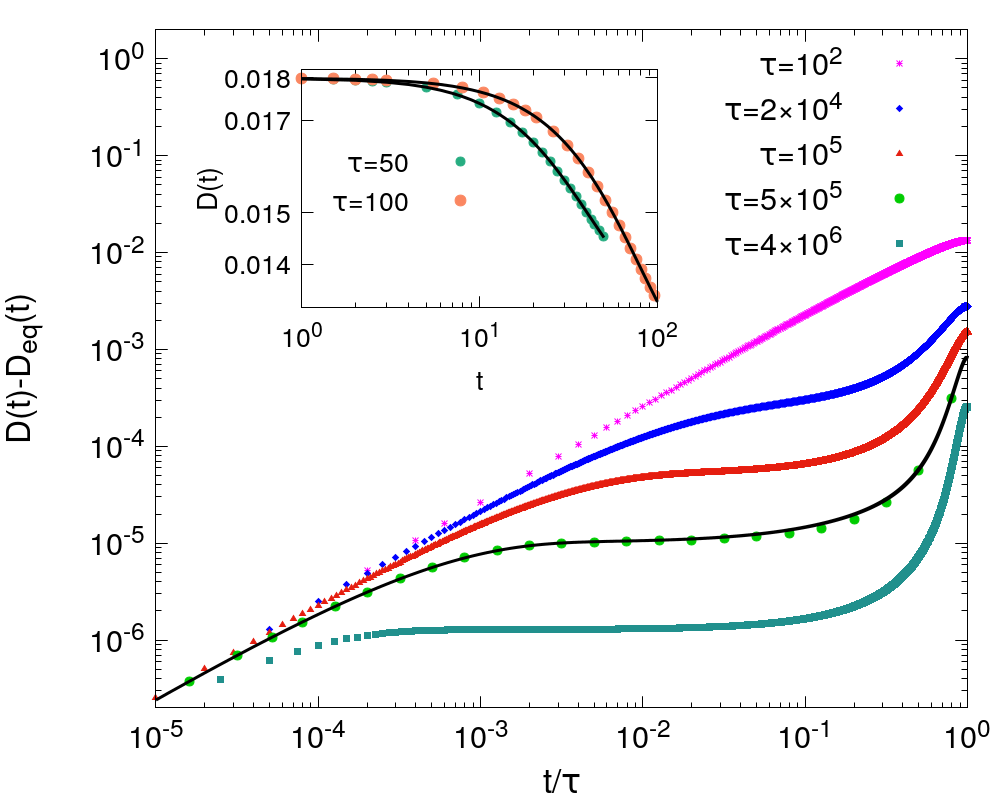}
         \caption{}
         \label{fig1a}
     \end{subfigure}
         \begin{subfigure}{0.49\textwidth}
         \centering
         \includegraphics[width=1.0\textwidth]{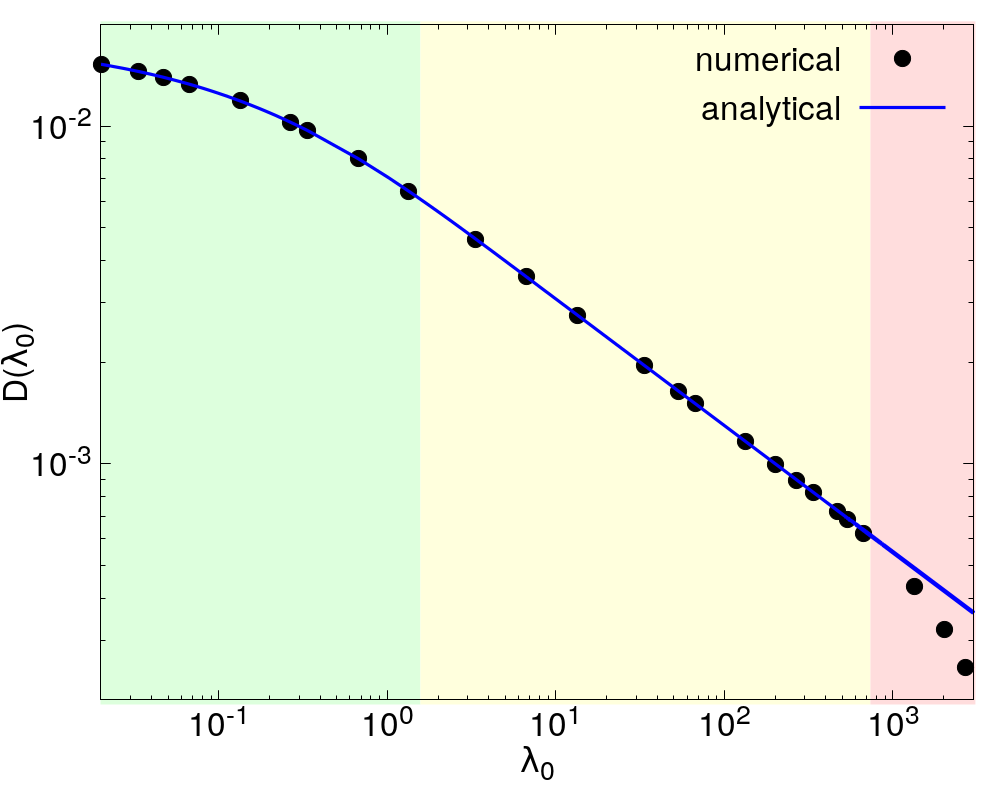}
        \caption{}
        \label{fig1b}
     \end{subfigure}
    \caption{Glauber Ising chain when the system is initially equilibrated to a low temperature $T_0$ and then slowly quenched to zero temperature: (a) The inset and main figure, respectively, show the dynamics of defect density and excess defect density for various quench times, and are obtained by numerically solving the exact equation (\ref{equn1}) (dots) which are compared for representative values of $\tau$ with (\ref{Gyx}) from low temperature theory (black solid lines).  (b) The figure shows the defect density at the end of the quench as a function of $\tau$ in three different regimes (represented by different colors) for a fixed $T_0$ and $L$ where, the line depicts the analytical solution (\ref{e14}). In these figures, the system size $L=2000$, $T_0=0.5$ and corresponding $\xi_0 \approx 27.3$, and $\alpha=3$ in the cooling protocol (\ref{equn2}).}       
\label{fig1}
\end{figure}

We first consider the situation when the system of size $L$ initially in the equilibrium state at a low temperature $T_0$ is  slowly quenched to zero temperature using the cooling protocol (\ref{equn2}). 
As the initial correlation length $\xi_0 \equiv \xi_{eq}(T_0)$ is large, a perturbation in the equilibrium state due to changing temperature will take time $\sim \xi_0^{z_{eq}}=\xi_0^{2}$ to relax (see Supplemental Material Fig.~\ref{fig2a}).  
Therefore if the quench time $ \tau \ll \xi_0^2$ (regime I), as shown in the inset of Fig. \ref{fig1a}, the defect density stays close to its equilibrium value at $T_0$ or  the excess defect density  increases with time (see Fig. \ref{fig1a} for $\tau=100$);  
thus the system can not enter the adiabatic phase and the defect density is not expected to follow the KZ scaling (\ref{KZIsing}) at late times. 
For larger quench times (regime II), as shown in Fig. \ref{fig1a} for $\tau > 100$, the excess defect density initially increases as $D(t)\approx D_{eq}(T_0)$ for $t \ll \xi_0^2$, but for  $\xi_0^2 \ll t \ll {\hat t}$, the system relaxes to the instantaneous temperature and the excess defect density remains constant; this adiabatic phase is followed by the KZ phase where the system can not keep up with the changing temperature due to diverging correlation length and the excess defect density increases for ${\hat t} \ll t < \tau$.
If the quench time is long enough that the finite-sized system can relax (regime III), that is, $\tau -{\hat t} \gtrsim L^{z_{eq}}$, the deviation of the defect density from its equilibrium value is essentially zero, see Fig. \ref{fig1a} for $\tau=4 \times 10^6$. 

Below we describe the dynamics of defect density quantitatively using the low temperature theory discussed in Appendix~\ref{smalleqG}; Fig. \ref{fig1a} shows a comparison of the defect density obtained by numerically solving the exact equation (\ref{equn1}) and the solution (\ref{Gyx}) from the low temperature theory for representative values of $\tau$, and we find a good agreement. In view of the dynamical phases discussed above, the mean defect density at the end of the quench falls in three distinct regimes that are shown in Fig. \ref{fig1b}. For an infinitely large system, using (\ref{I1final}) and (\ref{I2final}), we find that in regime I and II, the exact expression for $D(\tau)$ is given by
\bea
D(\tau) &=& \frac{1}{2\sqrt{\pi\tau}} \left[ \left(\frac{2\lambda_0}{\alpha+1} \right)^\frac{1}{2 \alpha + 2} \Gamma\left( \frac{2 \alpha + 1}{2 \alpha + 2}\right)+ \frac{1}{2 (\alpha +1)} E_{\frac{3+2 \alpha}{2 \alpha+2}}\left(\frac{2\lambda_0}{\alpha+1} \right) \right] \nonumber \\ 
&-& \frac{ e^{-\frac{2 \lambda_0}{\alpha +1}}}{2\sqrt{\pi \tau}} \left[1- e^{2 \lambda_0} \sqrt{2\pi \lambda_0} \;\text{erfc}\left( \sqrt{ 2\lambda_0}\right)\right]
\label{e14}
\eea
where $\lambda_0=\tau (1-\gamma_0) \sim \frac{\tau}{\xi_0^2}$ and $E_n(z)$ is the exponential integral function, and matches the numerical results shown in Fig.~\ref{fig1b}. We now discuss these regimes in detail:

\noindent {\bf Regime I:} for $\lambda_0 \ll 1$ or $\tau \ll \xi_0^2$, as the system does not get enough time to relax to the slowly changing temperature, it is always in a nonequilibrium state. 
From (\ref{app_I1approx}) and (\ref{I2e1}), we find that at short times 
\be
D_I(t) \approx \sqrt{\frac{1-\gamma_0}{2}} ~,~ t \ll (1-\gamma_0)^{-1}
\ee
so that the defect density remains close to its initial value, {\it viz.}, the equilibrium defect density $D_{eq}(T_0)$ given by (\ref{equildef}). If the quench time is not too small, the defect density evolves and decreases with time (see the inset of Fig.~\ref{fig1a}). At the end of the quench, from (\ref{e14}), we find that the defect density  is given by  
\be
D_I(\tau) =
\sqrt{\frac{1-\gamma_0}{2}} \left(1- \sqrt{\frac{2  \lambda_0}{\pi}} \frac{4 \alpha}{ 2 \alpha +1} \right)~,~\lambda_0 \ll 1
\label{e12}
\ee
The first factor on the RHS of the above equation is simply the equilibrium defect density as the system stays close to the initial state due to diverging correlation length and small quench time, and the second factor which depends on the details of the cooling protocol captures the reduction in the defect density from $D_{eq}(T_0)$ due to changing temperature.

\noindent {\bf Regime II:} For  $\xi_0^2 \ll \tau \ll L^{1/\beta}$, from (\ref{app_I1}) and (\ref{I2}),  we find that 
\bsn
{D_{II}(t) \approx} 
\sqrt{\frac{1-\gamma_0}{2}} ~&,~$0< t \ll  (1-\gamma_0)^{-1}$\\
\sqrt{\frac{(1-\gamma_0)(1-\frac{t}{\tau})^\alpha}{2}}~&,~ $(1-\gamma_0)^{-1}\ll t \ll  \frac{\tau}{2}$
\esn
which, on comparing with (\ref{equildef}), show that the defect density is close to its initial value at very short times and then enters the adiabatic phase. At later times ($t \gg {\hat t}$) where the system is in the KZ phase, the dynamics are described by (34) of  \cite{Priyanka_2021} for $\gamma_0=0$ and we do not discuss them here. 
But at the end of the quench, from (\ref{e14}) [or, alternatively, adapting the analyses  of \cite{Priyanka_2021} to nonzero $\gamma_0$], we obtain 
\be
D_\textrm{II}(\tau) = (1-\gamma_0)^{\frac{1}{2(1+ \alpha)}} D_\textrm{II}(\tau, \gamma_0=0) ~,~\lambda_0 \gg 1
\label{eq2}
\ee
where $D_\textrm{II}(\tau, \gamma_0=0)$ is given by (\ref{eq1}). As expected, the defect density at the end of the quench is smaller when the system is initially equilibrated to low temperatures than when one starts with high temperatures.

%

\noindent {\bf Regime III:} For $\tau \gg \xi_0^2$, in an infinitely large system, the defect density at the end of the quench is inversely proportional to the correlation length, $\xi({\hat t})$. But in a finite system, we expect that 
\be
  D_\textrm{III}(\tau) = \frac{1}{\xi({\hat t})}\; {\tilde F}\Big(\frac{\xi({\hat t})}{L}\Big)=\frac{1}{\tau^{\beta}} \; F\Big(\frac{\tau}{L^{1/\beta}}\Big) 
   \label{equ8}
\ee
where the scaling function $F(w)$ is a constant for $\tau \ll L^{1/\beta}$ (regime II) and decays rapidly for $\tau \gg L^{1/\beta}$ towards the equilibrium value ({\it viz.}, zero) [see Supplemental Material Fig.~\ref{fig2b}]. 
Na{\"i}vely, one may expect that the quench time  over which the finite system reaches the  equilibrium state scales as $L^{z_{eq}}=L^2$ but, as stated above, the system relaxes to equilibrium  if the quench time $\tau \sim L^{1/\beta}$. Thus the quench time in which the system reaches the equilibrium state is {\it non-universal}, and the scaling exponent $\frac{1}{\beta}=2 +\frac{2}{\alpha}$ for finite-time cooling is larger than that for instantaneous quench. 
Note, however, that (\ref{equ8}) assumes that $\xi_0 \ll L$ but, if the initial correlation length is as large as the system size, the system never reaches the adiabatic phase (see Supplemental Material Fig.~\ref{fig3a}) and the finite system relaxes to equilibrium when $\tau \sim L^2$. 

\subsubsection{Nonequilibrium initial condition: I}
\label{NIC1}

We now consider a situation where the finite-sized system is initially not in equilibrium at a low temperature $T_0$. 
Specifically, we assume that the system is in the equilibrium state at a high temperature $T_i \gg 1$ and then instantaneously cooled to a low temperature $T_0 \ll 1$; starting from the resulting nonequilibrium state, the system is slowly cooled from $T_0$ to zero using the cooling protocol (\ref{equn2}). 

\begin{figure}[t]
     \centering
    \begin{subfigure}{0.49\textwidth}
         \centering
         \includegraphics[width=1.0\textwidth]{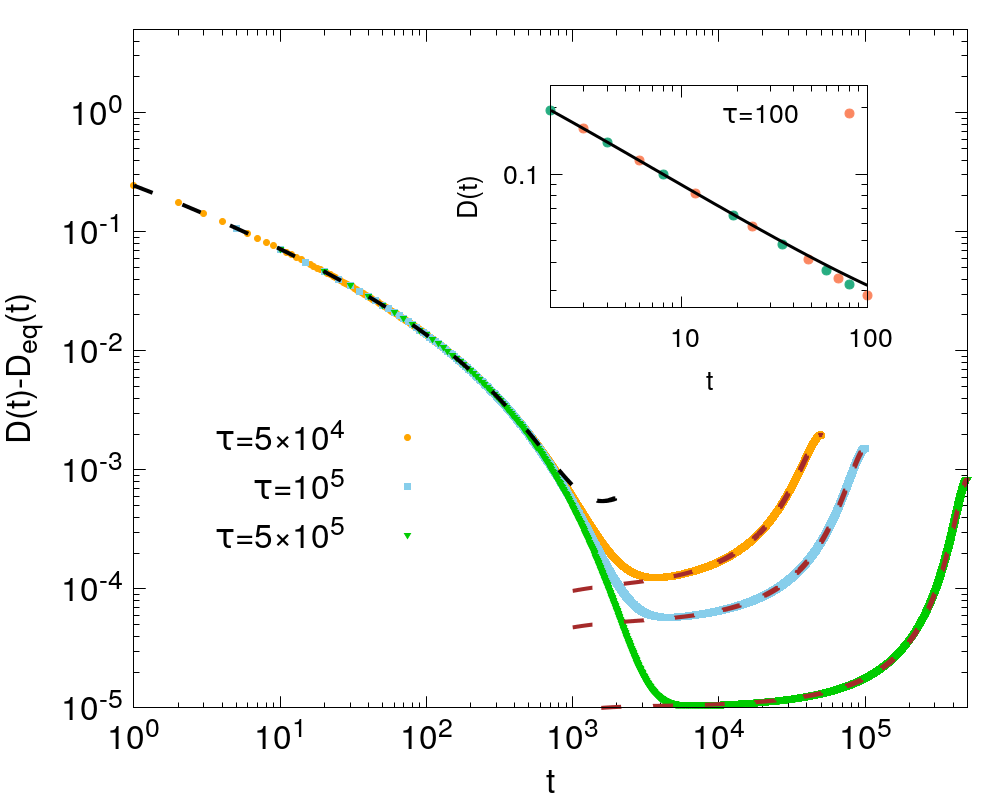}
         \caption{}
         \label{fig4a}
     \end{subfigure}
     \begin{subfigure}{0.49\textwidth}
         \centering
         \includegraphics[width=1.0\textwidth]{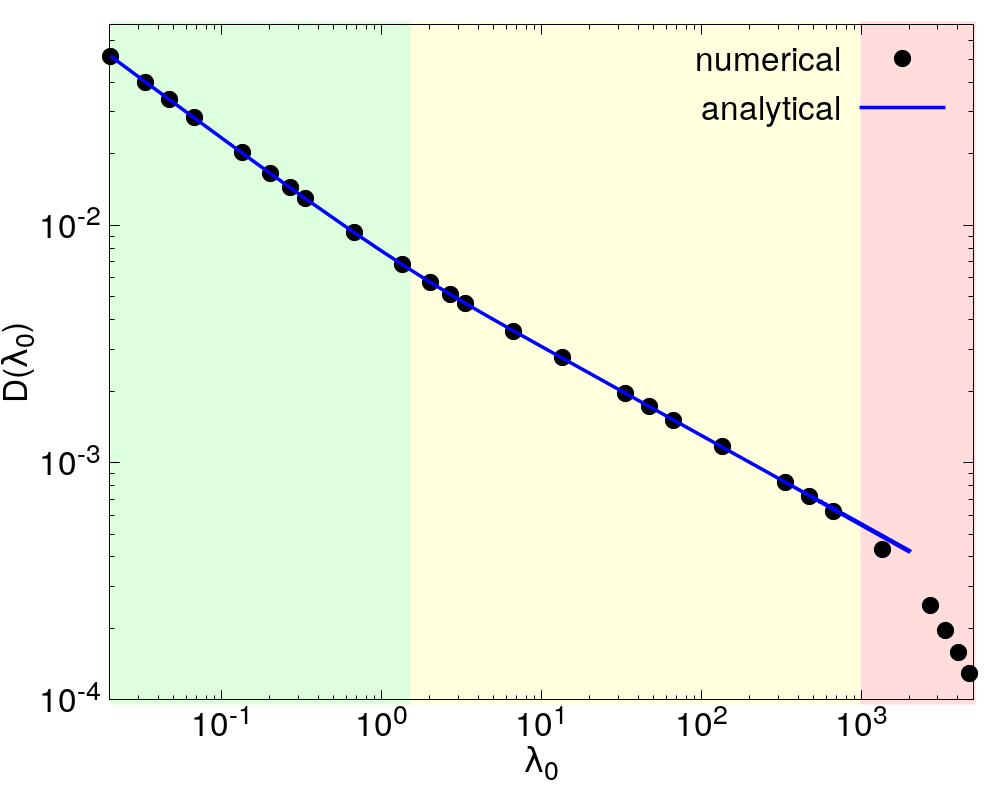}
        \caption{}
        \label{fig4b}
     \end{subfigure}
        \caption{Glauber Ising chain when the system is first instantaneously cooled from a high temperature $T_i$ to a low temperature $T_0$, and then slowly quenched to zero temperature: (a) The inset and main figure, respectively, show the dynamics of defect density and excess defect density for various quench times, and are obtained by numerically solving the exact equation (\ref{equn1}); in the inset, the green points show the comparison with (\ref{Gyx}) from low temperature theory. In these plots, the black line corresponds to the exact solution (\ref{eq3}) when the system is instantaneously cooled  to $T_0$ and the red dashed lines represent the excess defects obtained by numerically solving (\ref{equn1}) when the system is initially equilibrated to $T_0$. (b)
Figure shows the defect density at the end of the quench as a function of $\tau$ in three different regimes (represented by different colors) for a fixed $T_0$ and $L$ where, the line depicts the analytical solution (\ref{neqD2}). In these figures, the parameters are $L=2000$, $T_0=0.5$ and corresponding $\xi_0 \approx 27.3$, and $\alpha =3$ in the cooling protocol (\ref{equn2}).}
        \label{fig4}
\end{figure}

For $t \ll \tau$ where the variation in temperature can be neglected, the system behaves as if it is instantaneously quenched from $T_i$ to $T_0$ and undergoes coarsening dynamics; as Fig.~\ref{fig4a} shows, the dynamics of defect density under finite-time quench match those following a rapid quench from $T_i$ to $T_0$ until a time $t_0 \sim \xi_0^{z_{co}}=\xi_0^2$ as the system reaches the equilibrium state at $T_0$ on this time scale. 
If the quench time is small ($\tau \ll \xi_0^{z_{co}}$) so that the system can not equilibrate to $T_0$, it stays in the coarsening phase until the end of the quench (see the inset of Fig.~\ref{fig4a}) and the KZ scaling (\ref{KZIsing}) is not expected to hold. But if the quench time is long enough that the system can equilibrate to $T_0$, the dynamics are the same as discussed in the last subsection; in fact, as shown in Fig.~\ref{fig4a} for $\tau > 100$, the finite-time quench curve for nonequilibrium initial condition now matches the finite-time quench dynamics when the system is initially equilibrated to $T_0$ where, as discussed in Sec.~\ref{EIC}, the dynamics are in adiabatic phase ($t_0 \ll t \ll {\hat t}$) and KZ phase (${\hat t} \ll t < \tau$).

For an infinitely large system, from (\ref{I1final}) and (\ref{paraic}), we find that the mean defect density at the end of quench is given exactly by 
\be
\begin{aligned}
D(\tau) &= \frac{1}{2\sqrt{\pi \tau}} \left[ \left(\frac{2\lambda_0}{\alpha+1} \right)^\frac{1}{2 \alpha + 2} \Gamma \left( \frac{2 \alpha + 1}{2 \alpha + 2} \right)+ \frac{1}{2 (1+\alpha)} E_{\frac{3+2 \alpha}{2 \alpha+2}}\left(\frac{2\lambda_0}{\alpha+1} \right)\right] 
\end{aligned}
\label{neqD2}
\ee 
and matches the numerical results shown in Fig.~\ref{fig4b}. The dynamics of the defect density are quantitatively described below:

\noindent {\bf Regime I}: for $\tau \ll \xi_0^2$, the system is always in a nonequilibrium state and 
from (\ref{app_I1approx}) and (\ref{paraic}), we find that at short times
\be
D_I(t) \approx \frac{1}{2\sqrt{\pi t}} \left[1+ 2(1-\gamma_0) t \right]~,~ t \ll (1-\gamma_0)^{-1}
\ee
which matches the result (\ref{defs}) for rapid quench to $T_0$. However, for $t \lesssim \tau$ where the effect of changing temperature can not be neglected, the defect density curve starts diverging from the instantaneous cooling curve  (see the inset of Fig.~\ref{fig4a}). As a result, 
 the defect density at the end of the quench calculated from (\ref{neqD2}) is given by 
\be
D(\tau) = \frac{1}{2\sqrt{\pi \tau}} \left(1+\frac{2 \lambda_0}{2 \alpha ^2 +3 \alpha +1}\right)~,~ \lambda_0 \ll 1
\label{e13}
\ee
As explained for (\ref{eq1}), the above expression shows that the defect density at the end of the finite-time quench is larger than that for instantaneous quench to zero temperature. 
But, as the inset of Fig. \ref{fig4a} and a comparison between (\ref{defs}) and (\ref{e13}) show, it is smaller than when the system is instantaneously  quenched to $T_0$.

\noindent {\bf Regime II}: for $\xi_0^2 \ll \tau \ll L^{1/\beta}$, (\ref{app_I1}) and (\ref{paraic}) show that 
\bsn
{D_{II}(t) \approx}
 \frac{1}{2\sqrt{\pi t}} \left[1+2 (1-\gamma_0) t \right]~&,~ $0 \ll t \ll (1-\gamma_0)^{-1}$\\
 \sqrt{\frac{(1-\gamma_0)(1-\frac{t}{\tau})^\alpha}{2 }} ~&, $(1-\gamma_0)^{-1} \ll t \ll \frac{\tau}{2}$
\esn
and the defect density at the end of the quench is given by (\ref{eq2}).

\noindent {\bf Regime III}: For $\xi_0 \ll L, \tau \gg L^{1/\beta}$, the defect density has the same behavior as in regime III of Sec.~\ref{EIC}. But if $\xi_0 \sim L$, the defect density at the end of quench does not follow KZ scaling for any quench time as the system can not enter the adiabatic phase and instead, it decays according to the coarsening law until the system equilibrates (see Supplemental Material Fig.~\ref{fig3b}). 

\subsubsection{Nonequilibrium initial condition: II}
\label{NIC2}

\begin{figure}[t]
     \centering
    \begin{subfigure}{0.49\textwidth}
         \centering
         \includegraphics[width=1.0\textwidth]{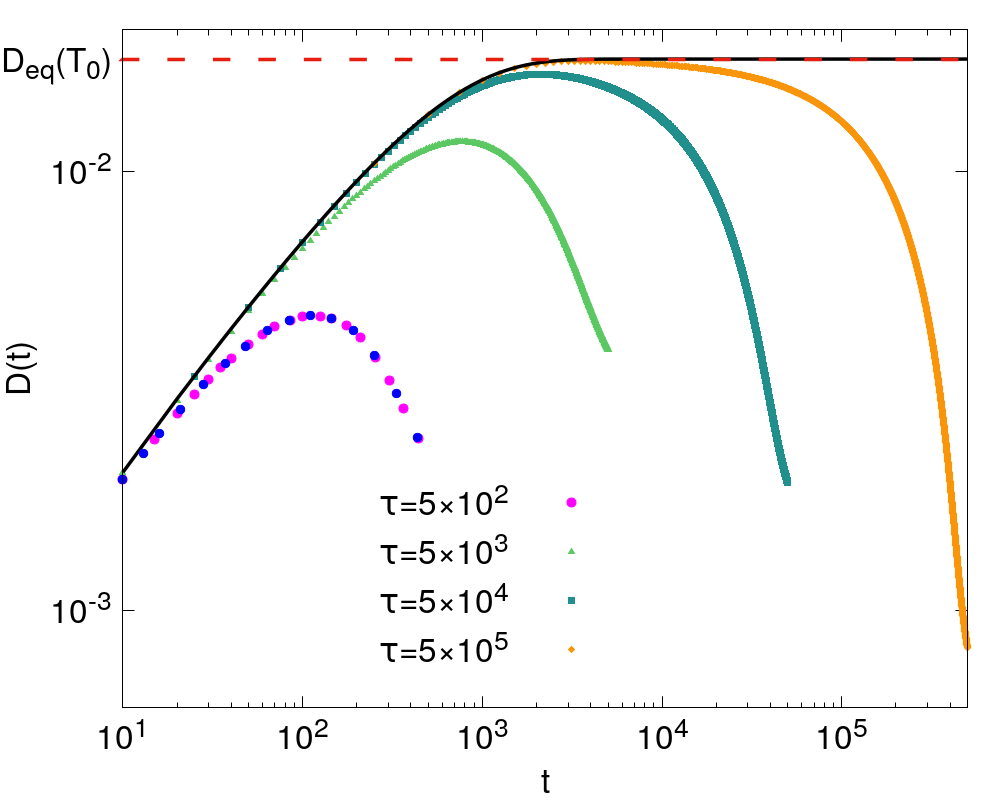}
         \caption{}
         \label{fig5a}
     \end{subfigure}
     \begin{subfigure}{0.49\textwidth}
         \centering
         \includegraphics[width=1.0\textwidth]{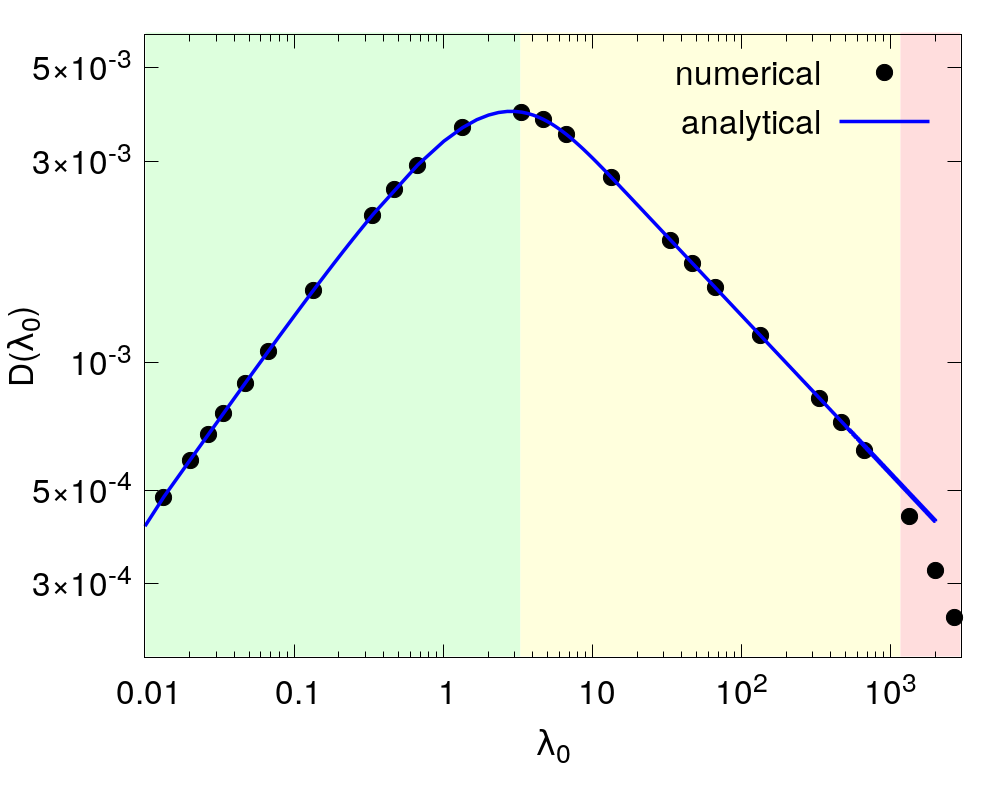}
        \caption{}
        \label{fig5b}
     \end{subfigure}
        \caption{Glauber Ising chain when the system in the equilibrium state at zero temperature is instantaneously heated to a low temperature $T_0$ and then slowly cooled to zero temperature: (a) The figure shows the dynamics of the defect density for various quench times obtained by numerically solving the exact equation (\ref{equn1}), and  the blue dots for $\tau=500$ show the comparison with (\ref{Gyx}) from low temperature theory. As the quench time increases, the system is able to reach the equilibrium state at $T_0$ as shown by the red dashed line while the black line corresponds to the  exact solution (\ref{eq3}) when the system is instantaneously heated  to $T_0$. (b) 
     The  figure shows the defect density at the end of the quench as a function of $\tau$ in three different regimes (represented by different colors) for a fixed $T_0$ and $L$ where, the line depicts the analytical solution (\ref{neqD1}). In these figures, the parameters are $L=2000$, $T_0=0.5$ and corresponding $\xi_0 \approx 27.3$, and $\alpha =3$ in the cooling protocol (\ref{equn2}).}
        \label{fig5}
\end{figure}

We now consider a situation where a finite-sized system in the critical state (that is, zero temperature) is instantaneously heated to a low temperature $T_0$ and then slowly cooled to zero temperature using the cooling protocol (\ref{equn2}). At short times $t \ll \tau$ where the effect of changing temperature can be neglected,  as in Sec.~\ref{NIC1}, the system behaves as if it is instantaneously heated from zero temperature to a finite temperature $T_0$; this is verified  in Fig. \ref{fig5a} where the dynamics of defect density under finite-time quench match with those following a rapid heating from $T_i=0$ to $T_0$ till a time $t_0\sim \xi_0^2$ as the system reaches the equilibrium state at $T_0$ on this time scale. As in Sec.~\ref{NIC1}, now depending on whether $\tau$ is smaller or larger than $\xi_0^2$, the defect density at the end of quench shows different scalings. 

From (\ref{I1final}) and (\ref{criticfinal}), we find that for this protocol, the mean defect density at the end of quench is given exactly by 
\be
\begin{aligned}
D(\tau) &=\frac{1}{2\sqrt{\pi \tau}} \left[ \left(\frac{2\lambda_0}{\alpha+1} \right)^\frac{1}{2 \alpha + 2} \Gamma \left( \frac{2 \alpha + 1}{2 \alpha + 2}\right)+ \frac{1}{2 (1+\alpha)} E_{\frac{3+2 \alpha}{2 \alpha+2}}\left(\frac{2\lambda_0}{\alpha+1} \right)- e^{-\frac{2\lambda_0}{\alpha+1}} \right] 
\end{aligned}
\label{neqD1}
\ee 
and matches the numerical results shown in Fig. \ref{fig5b}.

\noindent {\bf Regime I}: for $\tau \ll \xi_0^2$, from (\ref{app_I1approx}) and (\ref{critic1}), we find that at short times, the defect density increases as 
\be
D_I(t) \approx \frac{2 (1-\gamma_0) \sqrt{t}}{\sqrt{\pi }} ~,~t \ll \frac{\tau}{2}
\ee
which matches the result (\ref{defc}) for instantaneous heating to temperature $T_0$. For $t \lesssim \tau$, the defect density decreases so that a peak in $D(t)$ occurs at a time that scales linearly with quench time. Furthermore, from (\ref{neqD1}), we find that the defect density at the end of the quench is given by
\be
D(\tau)= \frac{1}{\sqrt{\pi \tau}} \left(\frac{2\lambda_0}{2\alpha+1}\right) ~,~\lambda_0 \ll 1
 \label{e15}
\ee 
which, as expected, approaches zero as $\alpha\to \infty$.

\noindent {\bf Regime II}: for $\tau\gg\xi_0^2$, the defect density initially increases until it reaches $D_{eq}(T_0)$ followed by the adiabatic phase so that from (\ref{app_I1}) and (\ref{app_critic}), we have 
\bsn
{D_{II}(t) \approx}
 \frac{2 (1-\gamma_0) \sqrt{t}}{\sqrt{\pi }} ~&,~ $0 \ll t \ll (1-\gamma_0)^{-1}$\\
 \sqrt{\frac{(1-\gamma_0)(1-\frac{t}{\tau})^\alpha}{2 }} ~&, $(1-\gamma_0)^{-1} \ll t \ll \frac{\tau}{2}$
\esn
which is followed by the KZ phase where the defect density at the end of quench is given by (\ref{eq2}).

\noindent {\bf Regime III}: For $\tau \gg L^{1/\beta}$, we obtain the same behavior as in regime III described in Sec.~\ref{EIC}  provided $\xi_0 \ll L$ otherwise the regime II is absent and the defect density increases as $\sqrt{\tau}$ until the finite system equilibrates (see Supplemental Material Fig.~\ref{fig3c}).
 
\subsection{Dynamics of auto-correlation function}

In the last section, we discussed the equal time spin-spin correlation function, and here we briefly consider the unequal time spin-spin correlation function 
\be
C_n(t,t_w) = \langle\sigma_i(t_w)\sigma_{i+n}(t)\rangle
\ee
where $t_w \leq t$ is the waiting time. Using the conditional probability $p(\sigma,t|\sigma',t_w)$ of finding an infinitely large system in the state $\sigma$ at time $t$, given that it was in state $\sigma'$ at time $t_w < t$, 
as for the equal time correlation function, we can write the differential equation for the unequal time spin-spin correlation function as
\be
\frac{\partial}{\partial t} C_n(t,t_w) = -C_n(t,t_w) + \frac{\gamma(t)}{2}[C_{n-1}(t,t_w)+C_{n+1}(t,t_w)]
\ee 
where $-\infty < n< \infty$ with the boundary conditions $C_{-\infty}(t)=0$ and $C_{\infty}(t)=0$ and the intial condition $C_n(t_w,t_w)=G_n(t_w)$.

\begin{figure}[t]
    \centering
    \includegraphics[width=0.85\textwidth]{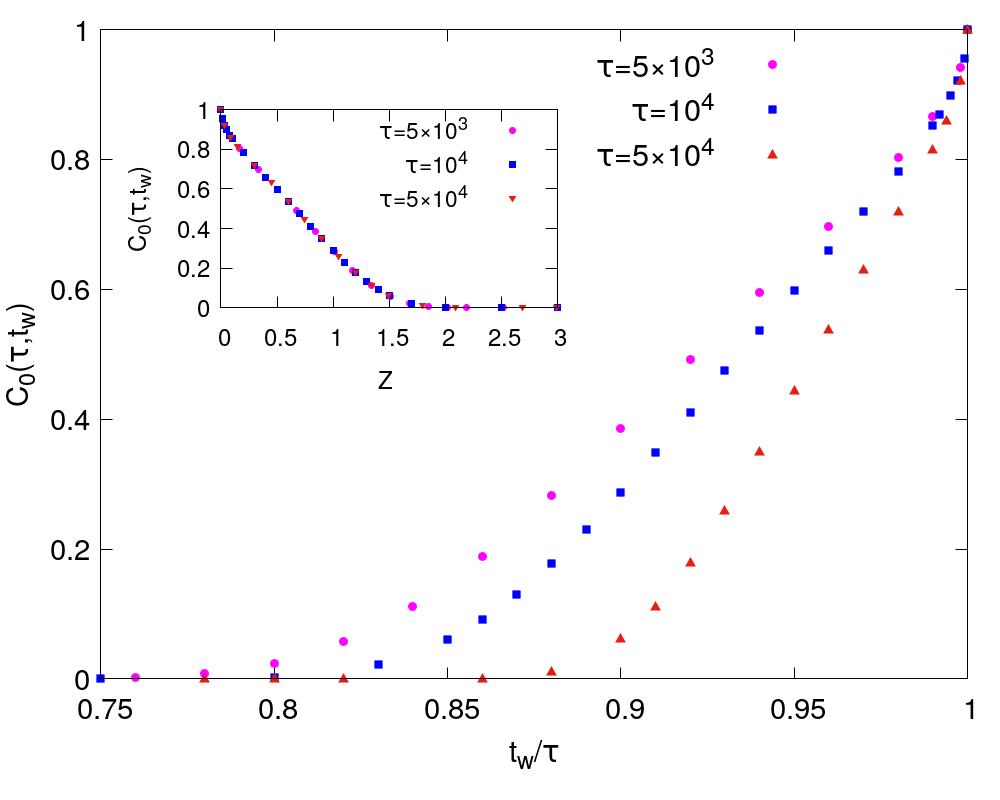}
    \caption{Glauber Ising chain when the system is quenched slowly from a high temperature to zero temperature: The main figure show the auto-correlation function between quench time $\tau$ and waiting time $t_w$, and the inset figure shows the data collapse according to KZ scaling ansatz (\ref{autocorr1}). The parameters are $L=2000$ and $\alpha=3$ in the cooling protocol (\ref{equn2}) with $\gamma_0=0$.}
    \label{autofig}
\end{figure}


When the system is instantaneously quenched from a high to a low temperature, the auto-correlation function, $C_0(t,t_w)$ decays as $\sqrt{\frac{t_w}{t-t_w}}, t \gg t_w$ \cite{brey1996low,prados1997aging}. But if the system is quenched at a finite rate, one expects that for $t_w < \hat{t}$, since the system is far from the critical point, a spin at the end of the quench is uncorrelated to its value at $t_w$ while for $t_w > \hat{t}$, the auto-correlation function is expected to increase to one as the system is close to the critical point where the correlation length is large. The auto-correlation function, $C_0(\tau,t_w)$ in Fig.~\ref{autofig} is indeed in agreement with these expectations, and shows that unlike the defect density which is almost frozen in the KZ phase (see, for example, inset of Fig. 3 in \cite{Priyanka_2021}), the auto-correlation function undergoes a large change over the same range of time. Furthermore, the inset of Fig.~\ref{autofig} shows that the data for different {quench and} waiting times can be collapsed onto a single curve if we assume the following scaling form 
\be
C_0(\tau, t_w) = {\cal C}\left(Z=\frac{\tau-t_w}{\tau-\hat{t}}\right)={\cal C}\left(Z=\frac{\tau-t_w}{\tau^{\frac{\alpha z_{eq}}{2+\alpha z_{eq}}}}\right)
\label{autocorr1}
\ee
{which is} in accordance with the KZ scaling (\ref{tscale}).


\section{Kawasaki Ising chain}

\subsection{Model}

In the last section, we have seen that both coarsening and stationary state dynamics play an important role in the finite-time quench dynamics. However, as the dynamic exponents for coarsening and stationary state dynamics are identical for the Glauber chain, below we consider the Kawasaki chain, for which these exponents are different, to understand how these affect the finite-time quench dynamics. 

Under  Kawasaki dynamics \cite{kawasaki1966diffusion}, the neighboring anti-parallel spins exchange  so that the magnetization remains strictly conserved. For time-dependent temperature, the master equation for the evolution of spin configurations can be written as 
\be
\begin{aligned}
     \frac{d}{dt}p(\sigma_{1},\ldots,\sigma_{i},\sigma_{i+1}, \ldots,\sigma_{L},t) &= \sum_{i=1}^L\Big[w(\sigma_{i+1} \leftrightarrow \sigma_{i},t )\; p(\sigma_{1},\ldots,\sigma_{i+1},\sigma_{i},\ldots,\sigma_{L},t) \\
&-   w(\sigma_{i}\leftrightarrow \sigma_{i+1},t )\;p(\sigma_{1},\ldots,\sigma_{i},\sigma_{i+1},\ldots,\sigma_{L},t) \Big]
     \label{equ3}
\end{aligned}
\ee
where the transition probability for the $i$th and $(i+1)${th} sites to exchange their spins is given by \cite{kawasaki1966diffusion}
\begin{equation}
 w(\sigma_{i} \leftrightarrow \sigma_{i+1},t )   = \left(1 - \frac{\gamma(t)}{2}(\sigma_{i-1}\sigma_{i}  + \sigma_{i+1}\sigma_{i+2})\right)\times \frac{1}{2}(1-\sigma_{i}\sigma_{i+1})
    \label{equ4}
\end{equation}
and, as in Sec.~\ref{GMo}, $\gamma(t) = \tanh{(2/T(t))}$ and its time-dependence is described by (\ref{equn2}). Thus in Kawasaki dynamics, the allowed moves are 
\begin{subequations}
\bea
  \uparrow\cdot\downarrow\cdot\uparrow\cdot\downarrow \hspace{0.5cm} &\stackrel{1+ \gamma}{\longrightarrow}& \hspace{0.5cm} \uparrow\uparrow\cdot\downarrow\downarrow \label{move1}\\
   \uparrow\uparrow\cdot\downarrow\downarrow \hspace{0.5cm} &\stackrel{1- \gamma}{\longrightarrow}& \hspace{0.5cm} \uparrow\cdot\downarrow\cdot\uparrow\cdot\downarrow \hspace{1cm} \hspace{1cm}  \label{move2}\\
    \uparrow\cdot\downarrow\cdot\uparrow\uparrow \hspace{0.5cm} &\stackrel{1}{\longrightarrow}& \hspace{0.5cm} \uparrow\uparrow\cdot\downarrow\cdot\downarrow \hspace{1cm} \hspace{1cm}  \label{move3}
    \label{Kdyn}
    \eea
\end{subequations}
where the dot represents the domain wall. While the domain walls  decrease and increase, respectively, via the moves (\ref{move1}) and (\ref{move2}), the defect density remains unchanged due to the diffusion move in (\ref{move3}). 

Before considering the finite-time quenches, we discuss the situation when the system is instantaneously quenched to zero temperature; due to conserved magnetization, at zero temperature, the system always gets stuck in a metastable state  which consists of domains of length two or more. Then, from (\ref{move2}), only the energy-raising  transition is possible  but that is not allowed at zero temperature (as the rate $1-\gamma=0$). Hence, the Kawasaki chain never reaches the equilibrium state of zero temperature. If now one quenches the system to a temperature slightly above zero, energy-raising events are allowed which can lead to domain growth (and hence equilibrium state) via diffusion and annihilation moves, but as the domain wall creation rate $1-\gamma \sim e^{-4/T}$ {is very small at low temperatures}, 
one can define a new time scale $t' = t e^{-4/T}$ so that the move (\ref{move2}) takes a finite time but other processes occur instantaneously. Using these accelerated dynamics \cite{BenNaim1998DomainND,godreche2004non}, it has been shown numerically and analytically that the domain length  grows as $t^{1/{z_{co}}}$ where the coarsening exponent $z_{co}=3$. In contrast, in the stationary state, the relaxation time grows as $\sim \xi_{eq}^{z_{eq}}$ where $z_{eq}=5$ \cite{zwerger1981critical}. Thus as a consequence of the conservation, the Kawasaki dynamics are slower than the Glauber dynamics where both these exponents are equal to two.  

\subsection{Dynamics of spin-spin correlation function}

Using the master equation (\ref{equ3}), we find that the evolution equation for the $n$-point correlation function, $\langle \sigma_{i_1} ... \sigma_{i_n} \rangle$ is not  closed as it depends on the $(n+2)$-point correlation functions resulting in an infinite hierarchy of equations \cite{cornell1991domain} and it does not seem possible to obtain analytical expressions for the defect density. Therefore, the results in the following subsections are obtained by simulating long Kawasaki chains in continuous time. 
In our simulations, an anti-parallel spin pair at site $i$ and $i+1$ exchange their value at time $t$ with probability $\frac{w(\sigma_i \leftrightarrow \sigma_{i+1},t)}{\sum_j w(\sigma_j \leftrightarrow \sigma_{j+1},t)}$, and 
the time $t+\delta t$ at which the next update occurs is found using that the increment time $\delta t$ is approximately exponentially-distributed with rate $\sum_j w(\sigma_j \leftrightarrow \sigma_{j+1},t)$. 
When there are large number of defects in the system, spin exchange occurs frequently, but when very few defects are left, the time between successive updates becomes large and therefore, close to zero temperature, the finite-time quench dynamics grind to a halt and it seems difficult to obtain accurate numerical results. However, for zero magnetization, we have measured the mean defect density in simulations by averaging over $5000$ independent runs which are discussed below. 

\subsubsection{Equilibrium initial condition}

We first consider the situation when the system initially equilibrated to a high temperature ($\gamma_0=0$) is slowly quenched to zero temperature according to  (\ref{equn2}). 
Using Monte Carlo simulations described above, we measured the defect density as a function of time for various quench times, and find that at short times, the excess defect density remains close to zero and then increases as the system falls out of equilibrium at time $\sim {\hat t}$ (data not shown). 
We expect that in the KZ phase (${\hat t} \ll t < \tau$), the excess defect density scales as \cite{Priyanka_2021}
\be
D(t)-D_{eq}(t)=\frac{1}{\tau^\beta}\; {\cal K}\left(Z=\frac{\tau -t}{\tau-\hat{t}} \right) =\frac{1}{\tau^\frac{\alpha}{2+5\alpha}} \;{\cal K}\left(Z=\frac{\tau-t}{\tau^{\frac{5 \alpha}{2 + 5 \alpha}}} \right) 
\label{KZK}
\ee
on using that the remaining time and the exponent $\beta$, respectively, are given by (\ref{tscale}) and (\ref{betadefn}), and the exponent $z_{eq}=5$ for these dynamics. 

\begin{figure}[t]
     \centering
    \begin{subfigure}{0.49\textwidth}
         \centering
         \includegraphics[width=1.0\textwidth]{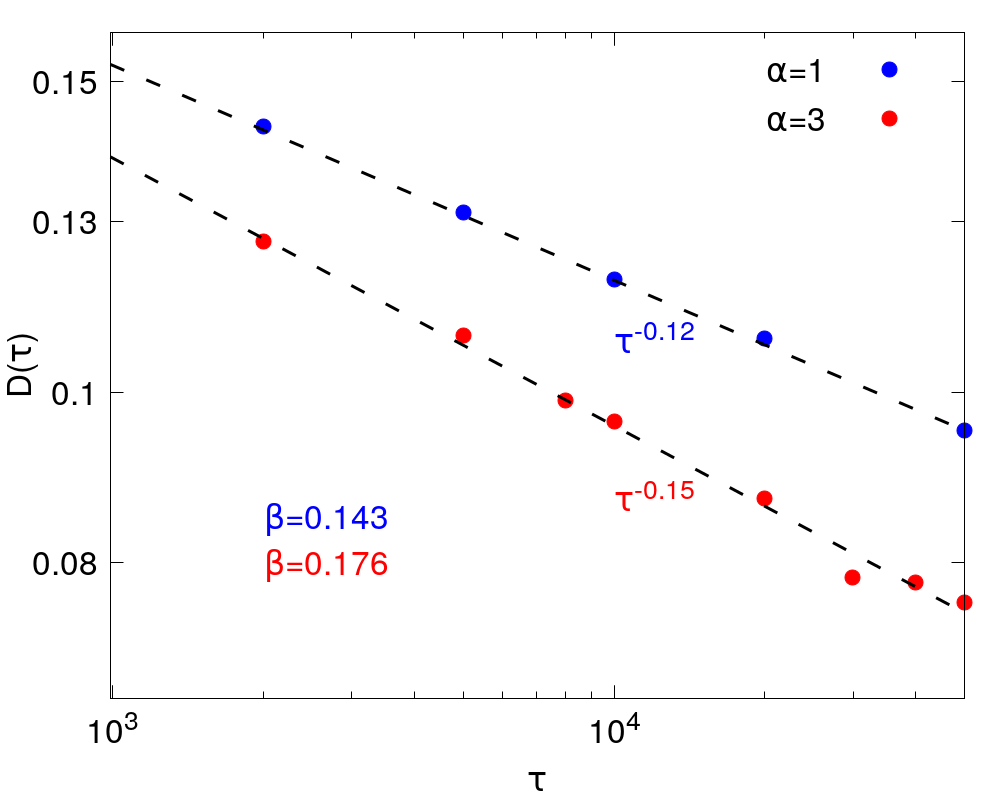}
         \caption{}
         \label{fig7a}
     \end{subfigure}
     \begin{subfigure}{0.49\textwidth}
         \centering
         \includegraphics[width=1.0\textwidth]{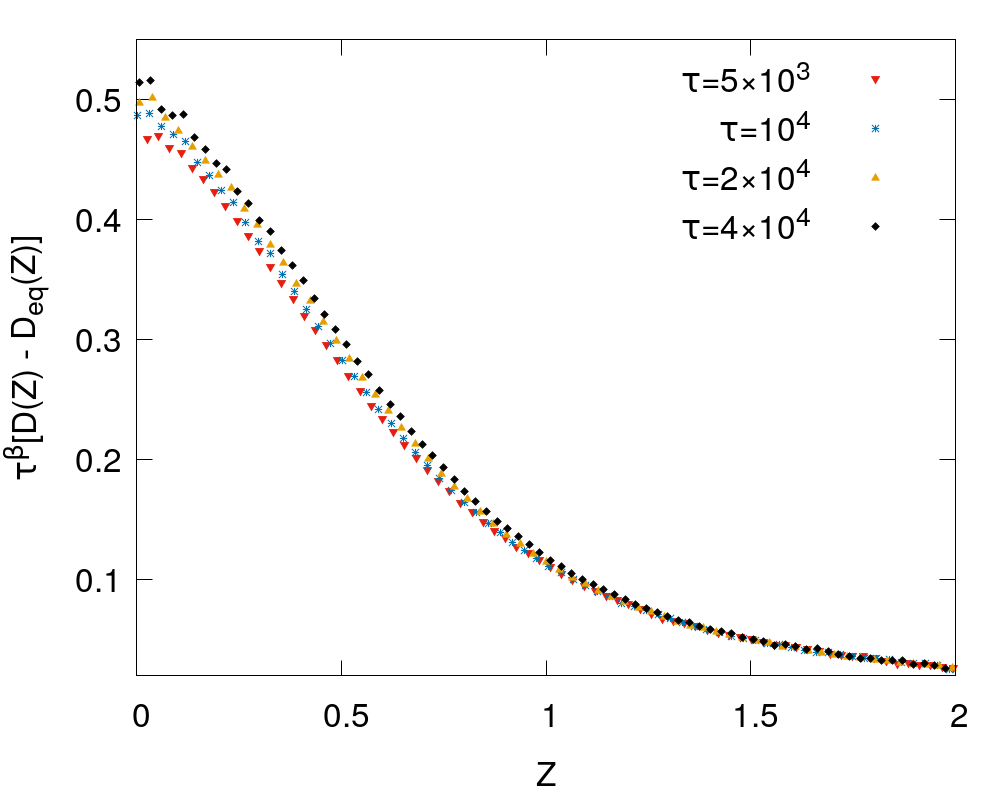}
         \caption{}
         \label{fig7b}
     \end{subfigure}
    \caption{Kawasaki Ising chain when the system initially equilibrated to a high temperature is slowly quenched to zero temperature: (a) The figure shows the  defect density at the end of the quench for two different $\alpha$ values in the cooling protocol (\ref{equn2}). Note that the exponents obtained are the best fits from numerical simulations which do not match exactly but are in close agreement with the KZ exponents quoted in the legend. (b) The figure shows the collapse of scaled excess defect density with the scaling variable $Z$ according to (\ref{KZK}) for $\alpha=3$ in the cooling protocol (\ref{equn2}). The system size $L=2000$ in both figures. }
\label{fig7}
\end{figure}

Fig. \ref{fig7a} shows that the exponent $\beta$ obtained from our simulations do not match exactly with (\ref{betadefn}) but their values are in fair agreement with the KZ predictions. Recently, the slowly quenched Kawasaki chain was studied numerically in \cite{kim2022nonequilibrium} where, for $\alpha=3$, the KZ exponent was found to be $\approx 0.163$ which is closer to the exact exponent, $\beta=3/17 \approx 0.176$ as compared to our best fit $0.15$ in Fig.~\ref{fig7a}, perhaps because much larger values of quench time ($\tau \gtrsim 10^7$) were used in \cite{kim2022nonequilibrium}. 
In Fig.~\ref{fig7b}, the scaling ansatz (\ref{KZK}) is tested and we find that  
a fairly good data collapse in the KZ phase is obtained so long as the temperature is not too close to zero; however, we also note that close to zero temperature  ($Z \approx 0$), the data collapse improves with increasing quench times. These results therefore support the KZ scaling following slow quench in the Kawasaki chain.

We also simulated the case where the system is initially equilibrated to a low temperature $T_0$ and then slowly quenched to zero temperature via the cooling protocol (\ref{equn2}). However, we were not able to check the scalings reliably due to the inability of the system to evolve at low temperatures as it gets stuck in the metastable states. But, as in Glauber chain (see Fig.~\ref{fig1b}), we expect that the defect density at the end of the quench has the following scaling form
\bsn
{D_{eq}(\tau)=\frac{1}{\xi_0} f_{eq} \left( \frac{\tau}{\xi_0^{z_{eq}}} \right) \propto}
\xi_0^{-1} ~&,~$\tau \ll \xi_0^{z_{eq}}$ \\
 \xi_0^{\beta z_{eq}-1} \tau^{-\beta} ~&,~$\tau \gg \xi_0^{z_{eq}}$
\label{Deq}
\esn
with $z_{eq}=5$. The behavior of the scaling function is deduced from the fact that 
at small quench times, the defect density remains essentially close to its initial value but for larger quench times, the system can enter the adiabatic phase leading to KZ scaling at the end of the quench.

\subsubsection{Nonequilibrium initial condition}

\begin{figure}[t]
    \centering
    \includegraphics[width=0.85\textwidth]{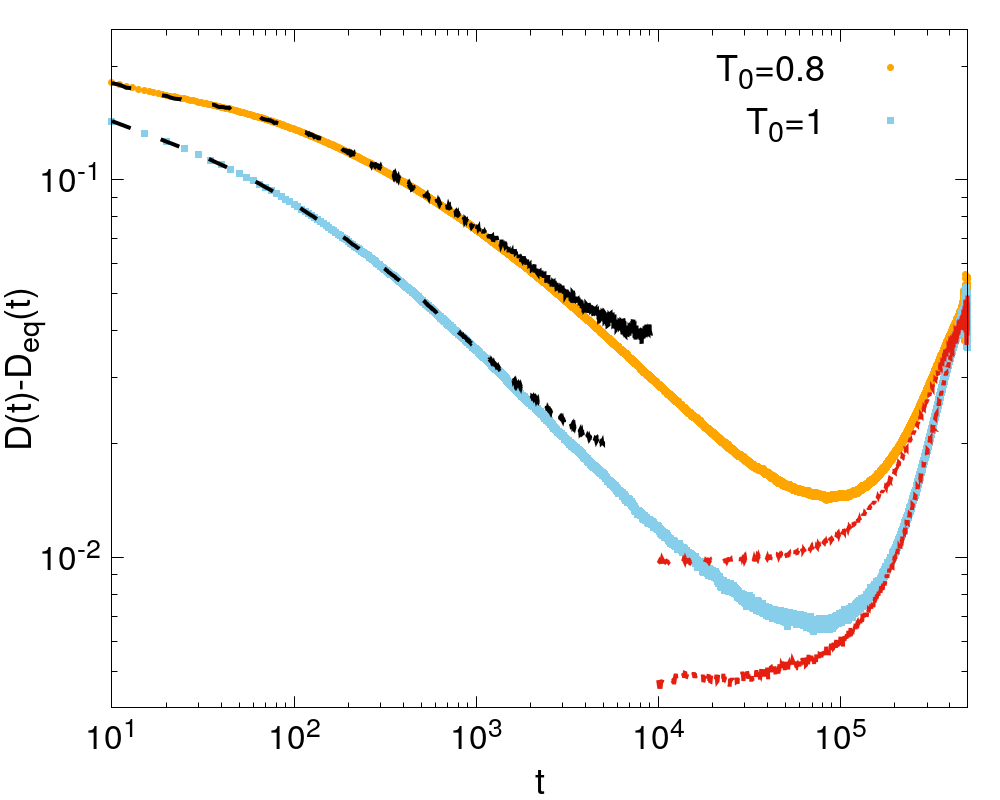}
    \caption{Kawasaki Ising chain when the system is first instantaneously cooled to a low temperature $T_0$, and then slowly cooled to zero temperature: The figure shows the dynamics of excess defect density for two initial temperatures $T_0$ at a fixed value of $\tau=5\times 10^{5}$. The black dashed line corresponds to the dynamics when the system is instantaneously cooled to $T_0$ and the red dashed lines represent the excess defects when the system is initially equilibrated to $T_0$. The parameters are $L=2000$ and $\alpha=3$ in the cooling protocol (\ref{equn2}).}
    \label{figK}
\end{figure}

We now consider the situation when the system initially at a high temperature $T_i$ is instantaneously cooled to a low temperature $T_0$ and then slowly quenched to zero temperature using cooling protocol (\ref{equn2}). For initial temperature $T_0 \ll 1$, as discussed in Sec.~\ref{NIC1} for Glauber chain, we expect that the defect density will decrease via coarsening and the system will reach the equilibrium state at temperature $T_0$ at time $t_0 \sim \xi_0^{z_{co}}$. However, as $T_0$ is small and the correlation length $\xi(t_0) \sim \xi_0$ is large, it will take time $\sim \xi_0^{z_{eq}}$ for a perturbation due to changing temperature to relax and we therefore expect that $D(t) \sim \xi_0^{-1}$ for $\xi_0^{z_{co}} \ll t \ll \xi_0^{z_{eq}}$. For $t \gg \xi_0^{z_{eq}}$, the system can enter the adiabatic phase followed by the KZ phase. We further conjecture that for an infinitely large system, at the end of the quench 
\bsn
{D_{neq}(\tau) \propto \label{DKneq}}
\tau^{-1/z_{co}} ~&,~$\tau \ll \xi_0^{z_{co}}$ \\
\xi_0^{-1}~&,~$\xi_0^{z_{co}} \ll \tau \ll \xi_0^{z_{eq}}$ \label{DKneq2}\\
\xi_0^{\beta z_{eq}-1} \tau^{-\beta} ~&,~$\tau \gg \xi_0^{z_{eq}}$
\esn
For the Glauber chain, as $z_{co}=z_{eq}$, the regime (\ref{DKneq2}) with constant defect density is not observed (see Fig.~\ref{fig4b}).

To test the above expectations, using Monte Carlo simulations, we measured the defect density for two initial temperatures as shown in Fig.~\ref{figK}. At short times $t\ll \tau$, the system behaves as if it is instantaneously quenched from a high temperature $T_i$ to $T_0$ and the finite-time quench curve matches the instantaneous quench dynamics. As the temperature is varying with time, the two curves start diverging at larger times, but the defects keep decreasing via coarsening. Note that unlike for the Glauber chain, here at a fixed $t$, on quenching the system from higher temperature results in lower number of defects because at higher $T_0$, the system does not get stuck in the metastable states and the spin updates occur more frequently (at least at short times) resulting in fewer defects. The excess defect density reaches a minimum when the system is in the adiabatic phase and then increases in the KZ phase where the finite-time quench curve matches the corresponding curve if the system started in equilibrium state at $T_0$. 

As discussed above, due to different dynamic exponents, we expect that the excess defect density will remain approximately constant (and close to zero) for $\xi_0^{z_{co}} \ll t \ll \xi_0^{z_{eq}}$; however, we do not observe this phase  in Fig.~\ref{figK} which, we believe, is because the initial  correlation length $\xi_0 \sim 10$ is quite small and the scaling regimes have not set in. Also, for the same reason, we have not been able to verify the scalings for the defect density at the end of the quench stated in (\ref{DKneq}). To observe these scalings and dynamical phases, we need to consider initial temperatures lower than those considered in Fig.~\ref{figK}. But even for $T_0=0.5, \xi_0 \sim 25$ (as considered in Glauber chain), the time $\delta t$ between successive updates is $\sim (1-\gamma_0)^{-1} \sim 600$ which gets longer as the temperature approaches zero and therefore, we need a better algorithm to capture the low temperature dynamics of the slowly quenched Kawasaki chain. 

\section{Discussion}

The Kibble-Zurek argument is a powerful and  general theory that predicts the density of defects when a classical or quantum system that exhibits second order phase transition is quenched from the disordered phase to critical region or ordered phase \cite{kibble1980some,zurek1996cosmological}. It assumes that if a system starts in an adiabatic phase, it will reach the KZ phase where the defect density decays as a  power-law with the quench time. In previous studies on finite-time quench dynamics in the Ising model, the system is assumed to be initially equilibrated to a high temperature and then cooled to the critical point \cite{brey1994dynamical, krapivsky2010slow,Priyanka_2021,godreche2022glauber,mayo2021distribution} or deep in the ordered phase \cite{biroli2010kibble,jain2016critical} at a finite rate, and one focuses on the defect density at the end of the quench (see, however, \cite{Priyanka_2021}). In contrast, here we studied the effect of initial conditions specified by the  initial state and initial temperature on the full dynamics till the end of the quench; we also elucidated how the system size affects the defect density.

We find that depending on the initial condition, besides the well known adiabatic and KZ phase, other dynamical phases such as coarsening are also possible; these are observed when the system starts in a nonequilibrium initial state which, to our knowledge, have not been considered in previous work. We formulated a low temperature theory for the Glauber Ising chain using which we obtained exact expressions (\ref{e14}), (\ref{neqD2}), (\ref{neqD1}) for the defect density at the end of quench for different initial conditions that are shown in Figs. \ref{fig1b}, \ref{fig4b}, \ref{fig5b}. 
  
As an application and extension of the scaling ideas developed for the Glauber chain, we also studied the Kawasaki Ising chain to understand the significance of different stationary state dynamic exponent and coarsening exponent. Since  the equations do not close for these dynamics, we performed Monte Carlo simulations but these simulations are also very hard as the system gets stuck in the metastable states at low temperatures. Therefore it remains to be seen if the scalings conjectured in (\ref{DKneq}) can be tested at low temperatures in the Kawasaki model or in some other model where the two dynamic exponents are quite different.


{\it Acknowledgements:} 
LJ would like to thank CSIR for fellowship.

\clearpage
\clearpage

\appendix
\setcounter{equation}{0}
\renewcommand{\thesection}{\Alph{section}}
\numberwithin{equation}{section}

\section{Rapid heating and cooling of Glauber chain}
\label{cnstT}

When an infinitely long Glauber chain is rapidly cooled or heated to a low temperature $T$ and then evolved at constant temperature $T$, the exact equation (\ref{equn1}) can be written as
\be
\frac{\partial G}{\partial x}=\frac{\partial^2 G}{\partial y^2}-2 G \label{cnstGeqn}
\ee
where, $x=t (1-\gamma), y=k \sqrt{1-\gamma}$ and the boundary conditions are $G(0,x)=1, G(\infty,x)=0$. The defect density (\ref{ddendef}) is then given by
\be
D(x)=\frac{1-G(\sqrt{1-\gamma},x)}{2} \stackrel{\gamma \to 1}{\to}-\frac{\sqrt{1-\gamma}}{2} \frac{\partial G}{\partial y}\bigg|_{y=0}
\ee
On taking the sine transform defined as $\tilde{G}(q,x)= \sqrt{\frac{2}{\pi}} \int_0^\infty dy\;\sin{ (q y)}\;G(y,x)$ of (\ref{cnstGeqn}), we obtain
\bea
{\tilde G}(q,x) = q \sqrt{\frac{2}{\pi}} \int_0^x dw e^{-(2+q^2) (x-w)}+{\tilde G}(q,0)e^{-(2+q^2)x} 
\eea
where ${\tilde G}(q,0)$ is the sine transform of the initial condition $G(y,0)$. 
The inverse sine transform then yields 
\bea
{G}(y,x) &=& {\frac{2}{\pi}} \int_0^\infty dq \sin(q y) \int_0^x dw q  e^{-(2+q^2) (x-w)}+\sqrt{\frac{2}{\pi}} \int_0^\infty dq \sin(q y){\tilde G}(q,0)e^{-(2+q^2)x} \\
&=& {\frac{1}{2\sqrt{\pi}}}  \int_0^x dw \frac{ y e^{\frac{-y^2}{4(x-w)}-2 (x-w)}}{(x-w)^{3/2}} +\sqrt{\frac{2}{\pi}} \int_0^\infty dq \sin(q y){\tilde G}(q,0)e^{-(2+q^2)x} 
\eea
where we have interchanged the order of integration in the first term. The above integrals correspond to $\alpha=0, \lambda_0=1$ of (\ref{app_I1defn}) and (\ref{app_I2defn}) which are analyzed in Appendix~\ref{defeqG}.

Alternatively, if we work with $H(y,x)=G_{eq}(y)-G(y,x)$ with homogeneous boundary conditions, $H(0,x)=H(\infty,x)=0$, we find that $H$ also obeys (\ref{cnstGeqn}) so that ${\tilde H}(q,x)={\tilde H}(q,0) e^{-(2+q^2)x}$. The defect density can be written as 
\bea
D(x) &=& \frac{\sqrt{1-\gamma}}{2} \left(\sqrt{2}+\sqrt{\frac{2}{\pi}} \int_0^\infty dq q {\tilde H}(q,x) \right)
\eea
For quench from high temperature to a low temperature $T$, using the initial condition ${\tilde H}(q,0)= \sqrt{\frac{2}{\pi}} \frac{q}{2+q^2}$, we obtain 
\bea
D(x) 
&=& \sqrt{\frac{1-\gamma}{2}}+\sqrt{\frac{1-\gamma}{\pi^2}} \left(\frac{\sqrt{\pi } e^{-2 x}}{2 \sqrt{x}}-\frac{\pi  \text{erfc}\left(\sqrt{2 x}\right)}{\sqrt{2}} \right) \\
&\stackrel{x \to 0}{\approx}&   \sqrt{\frac{1-\gamma}{\pi}} \left(\frac{1}{2 \sqrt{x}} + \sqrt{x}\right) \\
&=& \frac{1}{2 \sqrt{\pi t}}+(1-\gamma) \sqrt{\frac{t}{\pi}}~,~t \ll (1-\gamma)^{-1} \label{cnstcool}
\eea
Similarly, on heating the system from zero temperature to a low temperature $T$, as ${\tilde H}(q,0)= \sqrt{\frac{2}{\pi}} \frac{q}{2+q^2}- \sqrt{\frac{2}{\pi}} \frac{1}{q}$, we obtain
\bea
D(x) &=&\sqrt{\frac{1-\gamma}{2}} \text{erf}\left(\sqrt{2 x} \right) \\
&\stackrel{x \to 0}{\approx}&2 (1-\gamma) \sqrt{\frac{t}{\pi}}~,~t \ll (1-\gamma)^{-1} \label{cnstheat}
\eea

\section{Low temperature scaling theory for Glauber chain}
\label{smalleqG}


To describe the finite-time quench dynamics when an infinitely long Glauber chain is quenched from a low temperature $T_0$ to zero, we first rewrite the exact equation (\ref{equn1}) as 
\bea
     \frac{dG_k}{dt}
    &=&\gamma_0 (G_{k-1} + G_{k+1}-2 G_k)  -2G_k (1-\gamma(t)) \nonumber \\
    &+& (\gamma(t)-\gamma_0) (G_{k-1} + G_{k+1}-2 G_k) 
  \eea
The first term on the RHS of the above equation states that the dynamics are the same as when the system evolves at a time-independent, low temperature 
for which $\gamma_0 \to 1$ (see (\ref{equn1}) on replacing $\gamma(t)$ by $\gamma_0$) which is expected to be true for $t \ll \tau$ as the temporal variation of the temperature can be neglected. To take the effect of changing temperature into account,  
we consider the above equation in continuous space by writing $k'=k a$ where $a$ is the lattice spacing, and define $x=\frac{t}{\tau}$ to obtain
\bea
 \frac{\partial G(k',x)}{\partial x} &=&a^2 \tau \gamma_0 \frac{\partial^2 G(k',x)}{\partial k'^2} -2 \tau (1-\gamma_0) (1-x)^\alpha G(k',x) \nonumber \\
 &+& a^2 (1-\gamma_0) \tau \{1-(1-x)^\alpha\} \frac{\partial^2 G(k',x)}{\partial k'^2}
 \label{exeq}
\eea
For quenches from low temperatures, as $\gamma_0 \to 1$, we choose the lattice spacing $a=\sqrt{1-\gamma_0} \sim \xi_0^{-1}$. 
Then in the scaling limit $\tau \to \infty,  \gamma_0 \to 1$ such that $\lambda_0=\tau (1-\gamma_0)$  and $y=\frac{k'}{a \sqrt{\tau}}=\frac{k}{\sqrt{\tau}}$ are finite, as the last term on the RHS of the above equation is of order $1-\gamma_0 \sim \xi_0^{-2}$, it can be neglected and we finally arrive at 
\be
\frac{\partial G(y,x)}{\partial x} =\frac{\partial^2 G(y,x)}{\partial y^2}-2 \lambda_0 (1-x)^\alpha G(y,x)  \label{e3}
\ee

The above equation is subject to boundary conditions $G(0, x)=1$ and $G(\infty, x)=0$ (as the correlations are expected to vanish at large distances) and initial condition, $G(y,0)$. The exact solution for $G(y,x)$ can be obtained by taking the sine transform of (\ref{e3}) with respect to $y$ which yields the following first order differential equation 
\be
\frac{\partial \tilde{G}(q,x)}{\partial x} + \left(2\lambda_0 (1-x)^\alpha + q^2 \right)\tilde{G}(q,x) = \sqrt{\frac{2}{\pi}} q
\ee
where the sine transform is defined as $\tilde{G}(q,x)= \sqrt{\frac{2}{\pi}} \int_0^\infty dy\;\sin{ (q y)}\;G(y,x)$. Solving the above equation, we obtain
\bea
\tilde{G}(q,x) &=& \tilde{G}(q,0)\;e^{-q^2 x+\frac{2\lambda_0}{\alpha +1}\left((1-x)^{\alpha + 1}-1\right)} \nonumber \\ 
&+& e^{\frac{2\lambda_0}{\alpha +1}(1-x)^{\alpha + 1}-q^2 x} \int_0^x dw \;e^{q^2 w- \frac{2\lambda_0}{\alpha +1}(1-w)^{\alpha + 1}} \sqrt{\frac{2}{\pi}}q
\eea
where ${\tilde G}(q,0)$ is the sine transform of the initial condition $G(y,0)$. 
The inverse sine transform then yields 
\bea
G(y,x) &=& \frac{2}{\pi}\int_{0}^{\infty}dq\;\sin(qy)\;q\; e^{-q^2 x}\int_{0}^{x}dw\;e^{q^2 w -\frac{2\lambda_0}{\alpha+1}\big((1-w)^{\alpha+1}-(1-x)^{\alpha+1}\big)} \nonumber \\ 
&+& \sqrt{\frac{2}{\pi}}\int_{0}^{\infty}dq\;\sin(qy)\;\tilde{G}(q,0)\;e^{-q^2 x +\frac{2\lambda_0}{\alpha+1}\big((1-x)^{\alpha+1}-1\big)}
\label{e4}
\eea
and the defect density (\ref{ddendef}) is given by
\be
D(x)=\frac{G(0,x)-G(\tau^{-1/2},x)}{2} \stackrel{\tau \to \infty}{\to} -\frac{1}{2 \sqrt{\tau}} \frac{\partial G(y,x)}{\partial y} \bigg|_{y=0}
\label{defderiv}
\ee

\section{Defect density dynamics for Glauber chain}
\label{defeqG}

As shown in Appendix~\ref{smalleqG}, the two-point correlation function can be written as $G(y,x)=I_1+I_2$ 
where, $y=\frac{k}{\sqrt{\tau}}, x=\frac{t}{\tau}, \lambda_0=\tau (1-\gamma_0)$, 
\bea
I_1(y,x) &=& \frac{2}{\pi}\int_{0}^{\infty}dq\;q \sin(qy)\;e^{-q^2 x}\int_{0}^{x}dw\;e^{q^2 w -\frac{2\lambda_0}{\alpha+1}\big((1-w)^{\alpha+1}-(1-x)^{\alpha+1}\big)} \label{app_I1defn}\\
I_2(y,x) &=& \sqrt{\frac{2}{\pi}}\;e^{\frac{2\lambda_0}{\alpha+1}\big((1-x)^{\alpha+1}-1\big)}\int_{0}^{\infty}dq\;\textrm{sin}(qy)\;\tilde{G}(q,0)\;e^{-q^2 x} \label{app_I2defn}
\eea
and $\tilde{G}(q,0)$ is the sine transform of the initial condition $G(y,0)$. 

\noindent{{\bf Integral $I_1$:}} We first analyze the double integral $I_1$ which is independent of the initial condition. Interchanging the order of integration and on carrying out the integral over $q$ exactly in (\ref{app_I1defn}), we obtain
\bea
I_1(y,x) 
&=& \frac{y}{2\sqrt{\pi}}\int_{0}^{x}dw\;e^{-\frac{2\lambda_0}{\alpha+1}\big((1-w)^{\alpha+1}-(1-x)^{\alpha+1}\big)} \frac{e^{\frac{-y^2}{4(x-w)}}}{(x-w)^{3/2}} \label{I1next}\\
&=& \frac{1}{\sqrt{\pi}} \int_{\frac{y^2}{4x}}^{\infty}du\;\frac{e^{-u}}{\sqrt{u}}\;e^{-\frac{2\lambda_0}{\alpha+1}\big((1-x+\frac{y^2}{4u})^{\alpha+1}-(1-x)^{\alpha+1}\big)} 
\label{e5}
\eea

\noindent{\it Short time dynamics} ($x < 1/2$): The above integral is not exactly solvable but for $x < 1/2$, as $\frac{y^2}{4x} < \frac{y^2}{4 (1-x)} < u$, on expanding the integrand in the above expression in powers of $\frac{y^2}{4 (1-x) u}$ and retaining terms to leading order, we obtain
\bea
 I_1(y,x) &\stackrel{y\ll 1}{\approx}& \frac{1}{\sqrt{\pi}} \int_{\frac{y^2}{4x}}^{\infty}du\;\frac{e^{-u}}{\sqrt{u}}\;e^{-\lambda_0(1-x)^{\alpha}\frac{y^2}{2u}} \\
 &=& \frac{1}{2} \sum_{\epsilon=\pm 1} e^{\epsilon y\sqrt{2(1-x)^{\alpha}\lambda_0}}\textrm{erfc}\Big(\frac{y + 2\epsilon x\sqrt{2(1-x)^{\alpha}\lambda_0}}{2\sqrt{x}}\Big) 
\eea
Hence 
\bea
\frac{\partial I_1(y,x)}{\partial y}\bigg|_{y=0}&=& -\frac{ e^{-2 \lambda_0 x(1-x)^\alpha}}{\sqrt{\pi x}} +\frac{1}{2} \sqrt{2\lambda_0 (1-x)^\alpha} \sum_{\epsilon=\pm 1} \epsilon \;\textrm{erfc}\Big(\epsilon \sqrt{2x(1-x)^{\alpha}\lambda_0}\Big)\\
&\approx & -\frac{ e^{-2 \lambda_0 x}}{\sqrt{\pi x}} +\frac{1}{2} \sqrt{2\lambda_0 (1-x)^\alpha} \sum_{\epsilon=\pm 1} \epsilon \;\textrm{erfc}\Big(\epsilon \sqrt{2x\lambda_0}\Big)
 \label{e6}
   \eea
where we have written $x(1-x)^\alpha \approx x$ for $ x < 1/2$ 
which gives
\bsn 
{\frac{\partial I_1(y,x)}{\partial y}\bigg|_{y=0} \approx \label{app_I1}} 
 -\frac{1}{\sqrt{\pi x}}- \frac{2\lambda_0 \sqrt{x}}{\sqrt{\pi }} ~,~\lambda_0 x \ll 1 \label{app_I1approx} \\
-\sqrt{2\lambda_0(1-x)^\alpha} ~,~\lambda_0 x\gg 1 \label{app_I1approx2}
\esn
\noindent{\it At the end of quench} ($x=1$): For $1/2 < x < 1$, we have not been able to find a suitable approximation but we can obtain an expression for $I_1$ when $x=1$ (that is, at the end of the quench). For arbitrary $\lambda_0$ and $u_0=(\frac{2\lambda_0}{\alpha+1})^{\frac{1}{\alpha+1}} (\frac{y^2}{4})$, we rewrite (\ref{e5}) as
\bea
I_1(y,1)
&=& 1+\sqrt{\frac{u_0}{\pi}} \int_{0}^{\infty}du\;\frac{e^{-u u_0}}{\sqrt{u}}\;(e^{-(\frac{1}{u})^{\alpha+1}}-1) - \frac{1}{\sqrt{\pi}} \int_0^{\frac{y^2}{4}}du\;\frac{e^{-u}}{\sqrt{u}}\;e^{-(\frac{u_0}{u})^{\alpha+1}}\\
&\stackrel{y \to 0}{\approx}& 1+\sqrt{\frac{u_0}{\pi}} \int_{0}^{\infty}du\;\frac{e^{-(\frac{1}{u})^{\alpha+1}}-1}{\sqrt{u}}-\frac{1}{\sqrt{\pi}} \int_0^{\frac{y^2}{4}}du\;\frac{1}{\sqrt{u}}\;e^{-(\frac{u_0}{u})^{\alpha+1}} \\
&=& 1-\frac{y}{\sqrt{\pi}} \left(\frac{2\lambda_0}{\alpha+1} \right)^\frac{1}{2 \alpha + 2}\Gamma\left( \frac{2 \alpha + 1}{2 \alpha + 2} \right)-\frac{y}{2 (\alpha +1)\sqrt{\pi}} E_{\frac{3+2 \alpha}{2 \alpha+2}}\left(\frac{2\lambda_0}{\alpha+1} \right)
\eea
We therefore have
\be
\frac{\partial I_1(y,1)}{\partial y}\bigg|_{y=0}= -\frac{1}{\sqrt{\pi}} \left[ \left(\frac{2\lambda_0}{\alpha+1} \right)^\frac{1}{2 \alpha + 2} \Gamma( \frac{2 \alpha + 1}{2 \alpha + 2})+ \frac{1}{2 (\alpha +1)} E_{\frac{3+2 \alpha}{2 \alpha+2}}\left(\frac{2\lambda_0}{\alpha+1} \right)\right] \label{I1final}
\ee
where, {$E_n(z)=\int_1^\infty dw w^{-n} e^{-z w} $ is the exponential integral function}. We then obtain
\bsn
{\frac{\partial I_1(y,1)}{\partial y}\bigg|_{y=0}=}
- \frac{1}{\sqrt{\pi}} \left(1+\frac{2 \lambda_0}{2 \alpha ^2 +3 \alpha +1}\right) ,& $\lambda_0 \ll 1$ \\
-\frac{1}{\sqrt{\pi }}\left[\left(\frac{2\lambda_0}{\alpha +1}\right)^{\frac{1}{2\alpha +2}} \Gamma \left(\frac{2\alpha+1}{2\alpha +2}\right)+\frac{e^{-\frac{2\lambda_0}{\alpha +1}}}{4\lambda_0}\right] ,& $\lambda_0 \gg 1$
\esn


\noindent{{\bf Integral $I_2$:}} We now analyze the integral $I_2$ for different initial conditions:

\noindent{\it Equilibrium state at $T_0$:} For $G(k,0)=G_{k,eq} \approx e^{-k \sqrt{2(1-\gamma_0)}}=e^{-y \sqrt{2 \lambda_0}}$,
we obtain
\bea
I_2(y,x) = \frac{2}{\pi}\;e^{\frac{2\lambda_0}{\alpha+1}\big((1-x)^{\alpha+1}-1\big)}\int_{0}^{\infty}dq\;\textrm{sin}(qy)\;\frac{q}{q^2+2\lambda_0}\;e^{-q^2 x} 
\eea
which  gives
\bea
\frac{\partial I_2(y,x)}{\partial y}\bigg|_{y=0}&=&e^{-\frac{2\lambda_0}{\alpha+1}\big(1-(1-x)^{\alpha+1}\big)} \left[\frac{1- e^{2 \lambda_0 x} \sqrt{2 \pi \lambda_0 x} \;
   \text{erfc}\left(\sqrt{2 \lambda_0 x}\right)}{\sqrt{\pi x}} \right] \\ \label{equilic}
   &\stackrel{x< 1/2}{\approx} & e^{-2\lambda_0 x} \left[\frac{1- e^{2 \lambda_0 x} \sqrt{2 \pi \lambda_0 x} \;
   \text{erfc}\left(\sqrt{2 \lambda_0 x}\right)}{\sqrt{\pi x}} \right]
   \eea
and  therefore 
\bsn 
{\frac{\partial I_2(y,x)}{\partial y}\bigg|_{y=0}\approx \label{I2}}
-\sqrt{2\lambda_0} + \frac{1}{\sqrt{\pi x}}+\frac{2\lambda_0 \sqrt{x}}{\sqrt{\pi}} ~&,~ $\lambda_0 x \ll 1$ \label{I2e1} \\
{\cal O}(e^{-2\lambda_0 x}) ~&,~ $\lambda_0 x \gg 1$ \label{I2e2}
\esn
However, at the end of the quench ($x=1$)
\be
\frac{\partial I_2(y,1)}{\partial y}\bigg|_{y=0}=e^{-\frac{2\lambda_0}{\alpha+1}} \left[\frac{1- e^{2 \lambda_0} \sqrt{2 \pi \lambda_0} \;
   \text{erfc}\left(\sqrt{2 \lambda_0}\right)}{\sqrt{\pi}} \right] \\ \label{I2final}
   \ee


\noindent{\it Paramagnetic state at $T_0$:} If the system is in a paramagnetic state, $G(k,0)=\delta_{k,0}$ or $G(y,0) \sim \delta(y)$ but the sine transform ${\tilde G}(q,0)=0$ so that
\be
I_2(y,x)=0 \label{paraic}
\ee
at all times.


\noindent{\it Critical state at $T_0$:} If the system is initially in the critical state, the correlation function $G_k(0)=1$ for all $k$ and its sine transform $\tilde{G}(q,0)=  \sqrt{\frac{2}{\pi}}\frac{1}{q}$. Using this in (\ref{app_I2defn}), we obtain  
\bea
I_2(y,x) 
& =& e^{-\frac{2\lambda_0}{\alpha+1}\big(1-(1-x)^{\alpha+1}\big)} \textrm{erf}\left(\frac{y}{2\sqrt{x}}\right) \label{critic}
\eea
which yields
\bea
\frac{\partial I_2(y,x)}{\partial y}\bigg|_{y=0}&=&\frac{1}{\sqrt{\pi x}} e^{-\frac{2\lambda_0}{\alpha+1}\big(1-(1-x)^{\alpha+1}\big)} \\  \label{criticic}
&\stackrel{x<1/2}{\approx}& \frac{1}{\sqrt{\pi x}} e^{-2\lambda_0 x}
   \eea
and therefore
\bsn
{\frac{\partial I_2(y,x)}{\partial y}\bigg|_{y=0} \approx \label{app_critic}}
\frac{1}{\sqrt{\pi x}}-\frac{2\lambda_0 \sqrt{x}}{\sqrt{\pi}} ~&,~ $\lambda_0 x\ll 1$ \label{critic1}\\
{\cal O}(e^{-2\lambda_0 x}) ~ &,~ $\lambda_0 x\gg 1$ \label{critic2}
\esn
However, at the end of the quench ($x=1$)
\be
\frac{\partial I_2(y,1)}{\partial y}\bigg|_{y=0}=\frac{1}{\sqrt{\pi}} e^{-\frac{2\lambda_0}{\alpha+1}} \\ \label{criticfinal}
   \ee


\clearpage

\bibliography{References}

\clearpage

\appendix
\renewcommand{\appendixname}{Supplemental Material}
\renewcommand{\thesection}{\Alph{section}}
\setcounter{figure}{0}

\section{Glauber chain for equilibrium initial condition}
\begin{figure}[h!]
     \centering
    \begin{subfigure}{0.49\textwidth}
         \centering
         \includegraphics[width=1.0\textwidth]{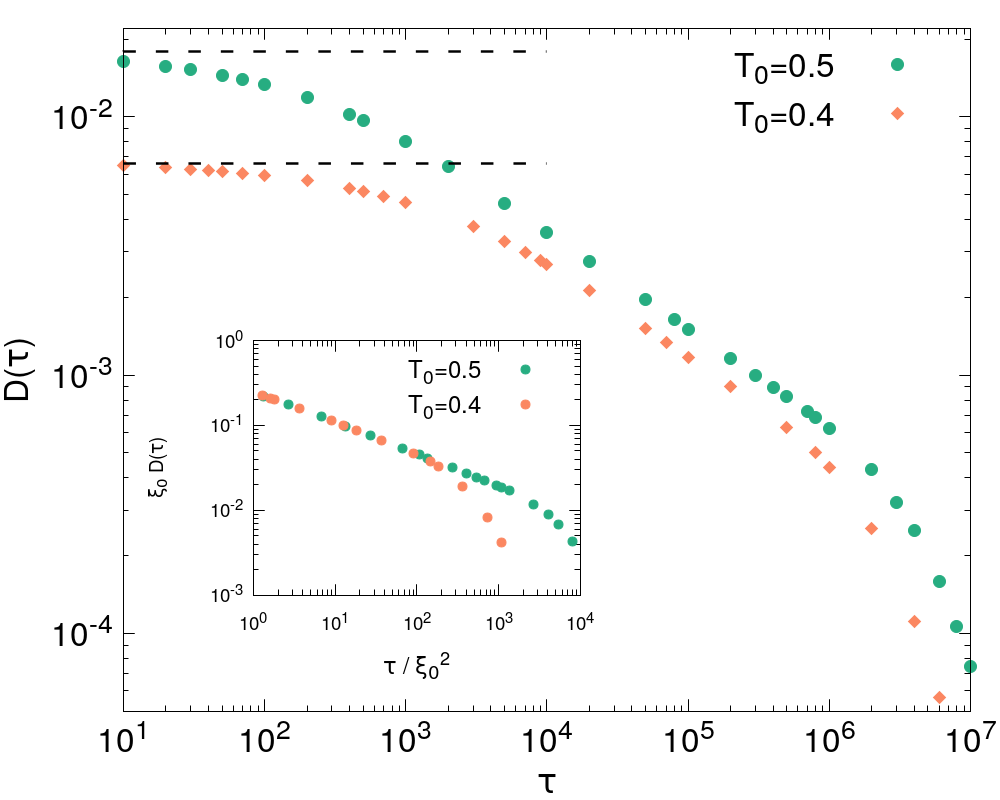}
         \caption{Effect of initial temperature}
         \label{fig2a}
     \end{subfigure}
     \begin{subfigure}{0.49\textwidth}
         \centering
         \includegraphics[width=1.0\textwidth]{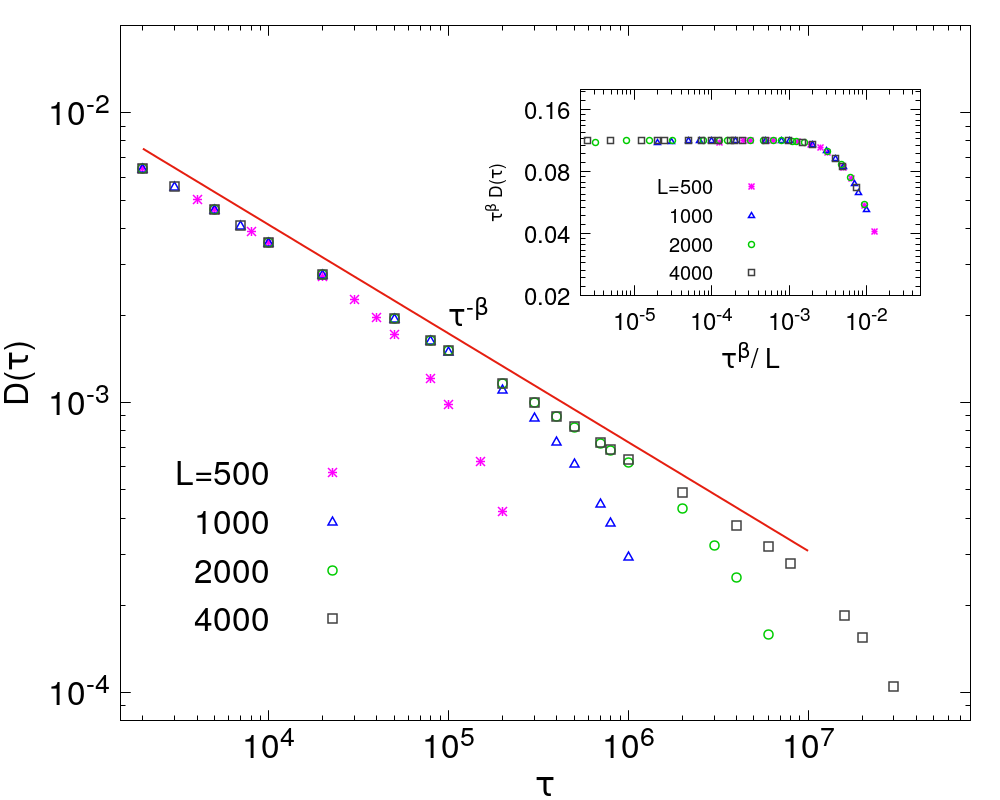}
         \caption{Effect of system size}
         \label{fig2b}
     \end{subfigure}
        \caption{Glauber Ising chain when the system initially equilibrated to a low temperature $T_0$ is slowly quenched to zero temperature: (a) The main figure shows the density of defects at the end of the quench for various quench times for two different $T_0$ values. The black dashed lines correspond to the equilibrium value (\ref{equildef}) at the respective initial temperatures. The inset figure shows the scaling collapse with $\tau \sim \xi_0^2$ in the regime I. (b) The main figure shows the effect of finite system size on the defect density at the end of the quench for various quench rates. The inset figure shows the scaling collapse in accordance with (\ref{equ8}). The parameters are $L=2000$, $T_0=0.5$ with $\xi_0 \approx 27.3$, and $\alpha=3$ in the cooling protocol (\ref{equn2}).}
        \label{SI_fig2}
\end{figure}

\clearpage

\section{Glauber chain for various initial conditions}

\begin{figure}[h!]
     \centering
    \begin{subfigure}{0.35\textwidth}
         \centering
         \includegraphics[width=1.0\textwidth]{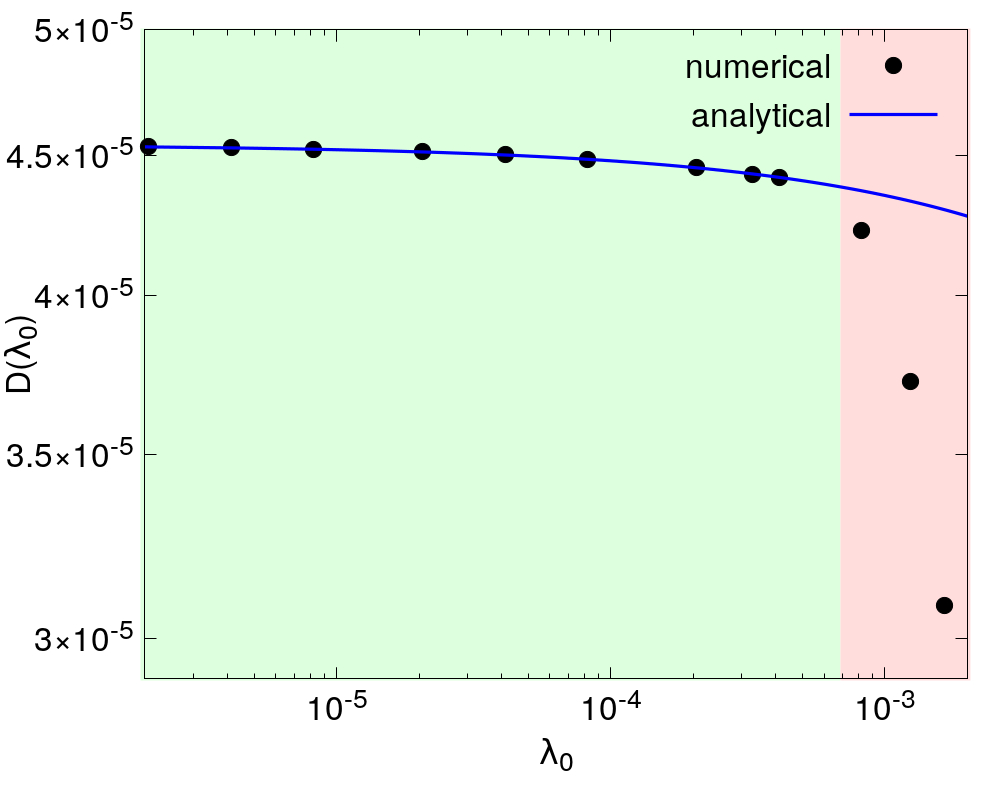}
         \caption{Equilibrium}
         \label{fig3a}
     \end{subfigure}
     \begin{subfigure}{0.35\textwidth}
         \centering
         \includegraphics[width=1.0\textwidth]{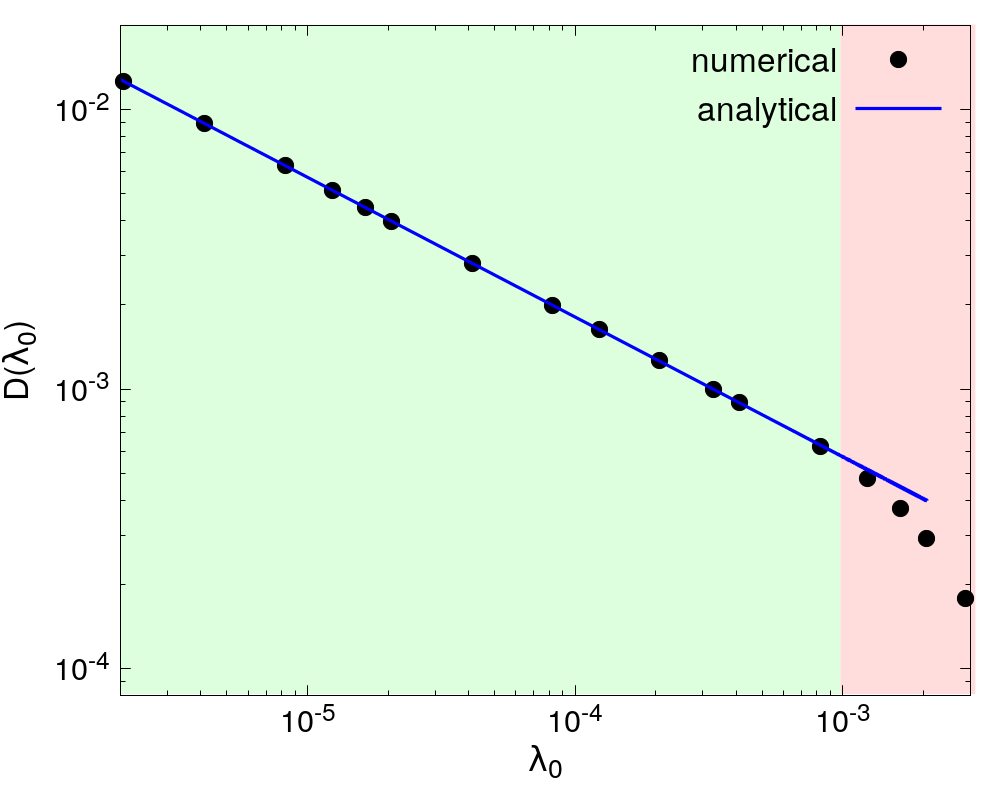}
         \caption{Nonequilibrium (cooling)}
         \label{fig3b}
     \end{subfigure}
     \begin{subfigure}{0.35\textwidth}
         \centering
         \includegraphics[width=1.0\textwidth]{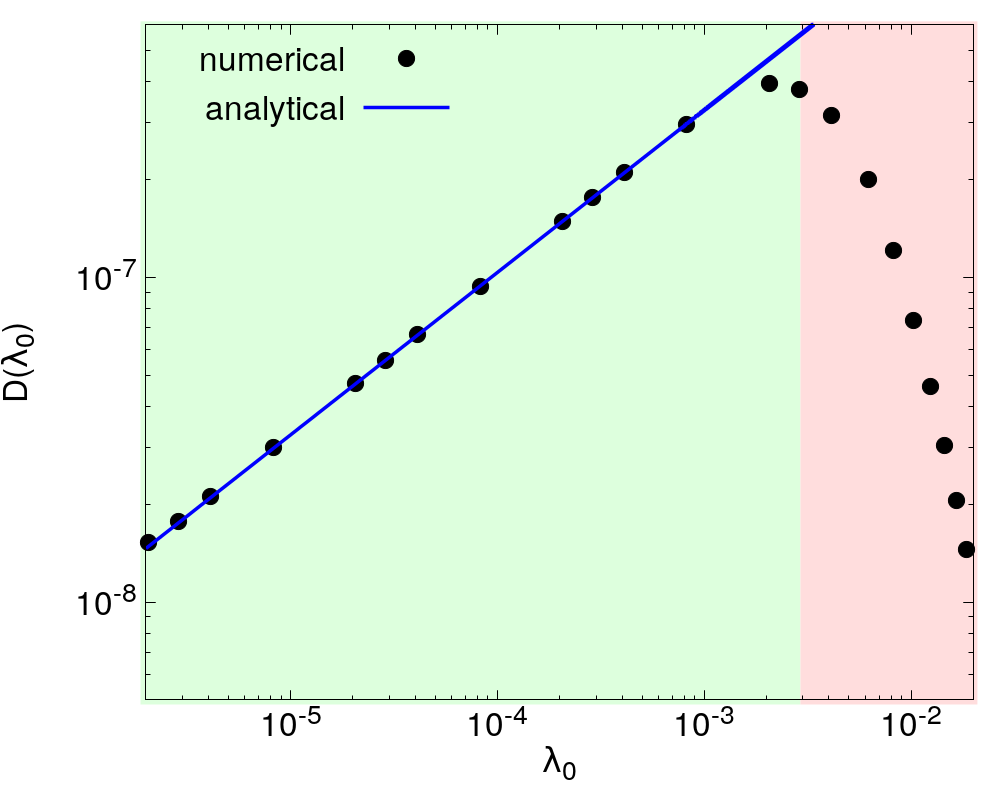}
         \caption{Nonequilibrium (heating)}
         \label{fig3c}
     \end{subfigure}
        \caption{Glauber Ising chain when quenched to zero temperature for various initial conditions and small system size: The figure (a) shows that the KZ phase is absent in the defect density at the end of the quench because the system is initially in equilibrium at a very low temperature where the correlation length $\xi_0$ is comparable to the system size. In (b) and (c) the system is in nonequilibrium state at $T_0$, the initial correlation length $\xi_i \sim {\cal O}(1)$ but $\xi_0 \sim L$ due to which system behaves as if it is instantaneously cooled or heated to zero temperature. The numerical data is obtained by solving the exact differential equation (\ref{equn1}) and the analytical data is obtained from (\ref{e14}), (\ref{neqD2}) and (\ref{neqD1}), respectively, for figures (a)-(c). The parameters are $L=2000$, $T_0=0.2$ and corresponding $\xi_{0}\approx 11000$, and $\alpha=3$ in the cooling protocol (\ref{equn2}).}
        \label{fig3}
\end{figure}

\end{document}